\documentclass[pdftex,twocolumn,epjc3,a4paper]{svjour3}          

\RequirePackage[T1]{fontenc}

\smartqed  

\RequirePackage{graphicx}
\RequirePackage{mathptmx}      
\RequirePackage{flushend}
\RequirePackage[numbers,sort&compress]{natbib}
\RequirePackage[colorlinks,citecolor=blue,urlcolor=blue,linkcolor=blue]{hyperref}

\journalname{Eur. Phys. J. C}

\usepackage[utf8]{inputenc}

\usepackage{amsmath}
\usepackage{amssymb}
\usepackage{graphicx}

\graphicspath{figs}
\usepackage{hyperref}
\usepackage{color}
\usepackage{commath}
\usepackage{url}
\usepackage{bm}
\usepackage{braket}
\usepackage{tikz}
\usetikzlibrary{positioning}
\usetikzlibrary{shapes.geometric, arrows}
\usepackage{url}
\usepackage{accents}
\usepackage{natbib}

\hypersetup{colorlinks=true,citecolor=blue, linkcolor=red, urlcolor=blue}

\newcommand{\inputnum}{1} 
\newcommand{\hiddennum}{4}  
\newcommand{\outputnum}{2}

\tikzstyle{startstop} = [rectangle, rounded corners, 
    text width=7em, minimum height=1cm,text centered, draw=black, fill=red!30]
\tikzstyle{io} = [trapezium, trapezium left angle=70, trapezium right angle=110, 
    text width=7em, minimum height=1cm, text centered, draw=black, fill=blue!30]
\tikzstyle{process} = [rectangle, 
    text width=7em, minimum height=1cm, text centered, draw=black, fill=orange!30]
\tikzstyle{decision} = [diamond, minimum height=1cm, text centered, text width=5.5em, node distance=3cm, draw=black, fill=green!30]
\tikzstyle{arrow} = [thick,->,>=stealth]

\begin{document}

\title{Neural network reconstruction of cosmology using the Pantheon compilation}

\author{Konstantinos F. Dialektopoulos\thanksref{e1,addr1a,addr1b}
\and 
Purba Mukherjee\thanksref{e2,addr2} 
\and 
Jackson Levi Said\thanksref{e3,addr3,addr4}
\and 
Jurgen Mifsud\thanksref{e4,addr4}
}

\thankstext{e1}{email: kdialekt@gmail.com}
\thankstext{e2}{email: purba16@gmail.com}
\thankstext{e3}{email: jackson.said@um.edu.mt}
\thankstext{e4}{email: jurgen.mifsud@um.edu.mt}

\institute{Department of Mathematics and Computer Science, Transilvania University of Brasov, Eroilor 29, Brasov, Romania\label{addr1a}
	\and
	Laboratory of Physics, Faculty of Engineering, Aristotle University of Thessaloniki, 54124 Thessaloniki, Greece\label{addr1b}
	\and
	Physics and Applied Mathematics Unit, Indian Statistical Institute, Kolkata - 700108, India\label{addr2}
	\and 
	Institute of Space Sciences and Astronomy, University of Malta, Malta, MSD 2080\label{addr3}
	\and
	Department of Physics, University of Malta, Malta, MSD 2080\label{addr4}
}

\date{Received: August 16, 2023 ~ / Accepted: October 8, 2023}

\maketitle

\begin{abstract}
In this work, we reconstruct the Hubble diagram using various data sets, including correlated ones, in Artificial Neural Networks (ANN). Using ReFANN, that was built for data sets with independent uncertainties, we expand it to include non-Guassian data points, as well as data sets with covariance matrices among others. Furthermore, we compare our results with the existing ones derived from Gaussian processes and we also perform null tests in order to test the validity of the concordance model of cosmology.
\end{abstract}





\section{Introduction} \label{sec:intro}

The standard model of cosmology is almost universally accepted as the concordance model for explaining cosmological observations \cite{Peebles:2002gy,Copeland:2006wr}. This is based on the incorporation of cold dark matter (CDM) to explain aspects of clustering \cite{Baudis:2016qwx,XENON:2018voc} while the late time accelerated expansion of the Universe \cite{Riess:1998cb,Perlmutter:1998np} is described through the action of a cosmological constant \cite{Mukhanov:991646}. While theoretical problems \cite{Weinberg:1988cp} of the cosmological constant description and the direct measurability of CDM \cite{LUX:2016ggv,Gaitskell:2004gd} have been in question for decades, the recent problems of cosmological tensions \cite{DiValentino:2020vhf,DiValentino:2020zio,DiValentino:2020vvd,Staicova:2021ajb,DiValentino:2021izs,Perivolaropoulos:2021jda,DiValentino:2022oon,SajjadAthar:2021prg} have brought into question the predictability of $\Lambda$CDM concordance model.

The cosmological tensions issue is most pronounced with the Hubble constant tension, which has shown a growing discrepancy between direct and indirect determinations of the $H_0$ parameter \cite{Abdalla:2022yfr}. The indirect approaches rely on assuming a $\Lambda$CDM cosmology \cite{Poulin:2023lkg} which is part of the reason why this model is being possibly reconsidered as the standard model of cosmology. In terms of indirect measurements, the latest reported values from the Planck and ACT collaborations are respectively $H_0^{\rm P18} = 67.4 \pm 0.5$ ${\rm km\, s}^{-1} {\rm Mpc}^{-1}$ \cite{Aghanim:2018eyx} and $H_0^{\rm ACT-DR4} = 67.9 \pm 1.5$ ${\rm km\, s}^{-1} {\rm Mpc}^{-1}$ \cite{ACT:2020gnv}, which point to a generically lower Hubble constant. On the other end of the spectrum, direct measurements of the Hubble constant have come from various different phenomenological sources. The strongest determination of the constant has come from the SH0ES team who have determined a best value of $H_0^{\rm R20} = 73.2 \pm 1.3$ ${\rm km\, s}^{-1} {\rm Mpc}^{-1}$ \cite{Riess:2020fzl}. This is based on observations of Type Ia Supernovae (SN-Ia) that are calibrated using Cepheid stars in their host galaxies. In this spirit, strong lensing measurements by quasar systems has also produced a consistent direct result of $H_0^{\rm HW} = 73.3^{+1.7}_{-1.8}$ ${\rm km\, s}^{-1} {\rm Mpc}^{-1}$ which is due to the H0LiCOW Collaboration \cite{Wong:2019kwg}. On the other hand, there is a direct result using the Tip of the Red Giant Branch (TRGB) technique which results in a lower value of the Hubble constant which gives $H_0^{\rm F20} = 69.8 \pm 1.9 \,{\rm km\, s}^{-1} {\rm Mpc}^{-1}$ \cite{Freedman:2020dne}. While systematics feature in every experiment, the Hubble tension appears to appear in several independent surveys and has now been present in several studies in the literature for some years.

The community has responded in several ways to this pressing problem. While work on understanding whether systematics may be the source of this tension will be ongoing for years to come, there is a growing body of work that is considering modifications to our standard picture of cosmology. The Hubble tension has been confronted with several interesting approaches in the literature including modifications to early Universe dark energy \cite{Poulin:2023lkg}, as well as the neutrino sector \cite{DiValentino:2021imh}, and renewed interest in modifications to gravitational models \cite{Addazi:2021xuf,CANTATA:2021ktz,Cai:2019bdh,Ren:2022aeo,Bernardo:2021qhu,Briffa:2020qli,LeviSaid:2021yat}. These approaches all offer interesting paths to new physics either through revisiting the foundations of cosmological models or by adding unknown components to the cosmological framework. However, many of these models are degenerate with each other in terms of current observational approaches which may require a new way of investigating new physics in the observational sector. One such approach is to consider the class of so-called model-independent methods. In this work, we aim to extend the current implementation of \textit{artificial neural networks} (ANN) \cite{10.2307/j.ctt4cgbdj} in terms of the Hubble diagram so that there will eventually be a way to perform reconstruction of cosmological models.

Through ANNs, real-world observational data can be used for undertaking reconstructions and inferences that are independent of any underlying physical models. They are also free of many of the statistical assumptions that appear in many of the other techniques. In this work, we reconstruct the Hubble diagram from various combined data sets where we fully incorporate the information in the data, specifically the covariance matrix. We do this by building on \texttt{ReFANN}\footnote{\url{https://github.com/Guo-Jian-Wang/refann}} \cite{Wang:2019vxv} which was originally designed for reconstructing the Hubble diagram for data sets with independent uncertainties, based on \texttt{PyTorch}\footnote{\url{https://pytorch.org/docs/master/index.html}}. We ran this code on GPUs which significantly reduced the computational time as compared with CPU runs. In Sec. \ref{sec:data}, we briefly introduce the data sets and discuss the reconstruction methodology adopted. We show the outputs for these analyses in Sec.~\ref{sec:ANN_results}. We compared and contrast our ANN outputs against their GP analogues in Sec.~\ref{sec:GP}. The null tests for these outputs are performed in Sec.~\ref{sec:null_tests}, while in Sec.~\ref{sec:conclusion} we discuss our main results and make some concluding remarks.

\section{Observational data sets \& Methodology} \label{sec:data}

In this part of the work we present the reconstruction methods used with a particular emphasis on ANNs and their architecture. We also discuss the data sets under investigation together with the priors used from the literature.

\subsection{Methodology}

The most popular approach to using model-independent techniques to study cosmology is through \textit{Gaussian Processes} (GP) \cite{10.5555/1162254} since they offer an integrated way to produce cosmological parameters together with their associated uncertainties. GP is based on a covariance function, or kernel, that characterizes the relationship between pairs of data points in a distribution. The kernel is functionally dependent on non-physical hyperparameters which can be fit using ordinary methods. The literature contains numerous works based on using this approach to reconstructing cosmological parameters \cite{Busti:2014aoa,Busti:2014dua,Seikel:2013fda,Bernardo:2021mfs,Yahya:2013xma,2012JCAP...06..036S,Shafieloo:2012ht,Benisty:2020kdt,Benisty:2022psx,Bernardo:2022pyz,Escamilla-Rivera:2021rbe,Bernardo:2021cxi,Mukherjee:2021epjc,Mukherjee:2022pdu}. Most recently, GPs have been used to reconstruct cosmological models \cite{Cai:2019bdh,Bernardo:2021qhu,Ren:2022aeo,Briffa:2020qli,LeviSaid:2021yat} from a foundational perspective. However, GP suffer from two major drawbacks, namely (i) they have an overfitting issue for low redshifts which can artificially constrain the Hubble constant at the level of its uncertainties; (ii) there is an over-reliance on the choice of kernel which may affect the profile of the reconstructed parameters.

An alternative approach to reconstructed cosmological parameters is through ANNs, which also open the way to the use of more complex data such as non-Gaussian data points and correlated data sets. Here, artificial neurons are modeled to mimic their biological counterpart, which are then organized into layers through which input signals are transformed into output signals. One example that this is formulated is input redshifts giving Hubble parameter and uncertainty outputs \cite{aggarwal2018neural,Wang:2020sxl,Gomez-Vargas:2021zyl}. An ANN is generally composed of a huge number of neurons that undergo training to optimize their associated hyperparameter values. A recent study in which this is performed is Ref.~\cite{Wang:2019vxv} which was further studied in Ref.~\cite{Dialektopoulos:2021wde} using null tests. Now, GP are a very attractive as an approach because they organically give higher order derivatives of their reconstructed function, and given that most cosmological models include such derivatives, they enter into the range of models that can be reconstructed in this way. In the recent work Ref.~\cite{Mukherjee:2022yyq}, the Hubble diagram ANN reconstruction method was extended to higher order derivatives using a Monte Carlo approach. This has opened the way for performing reconstructions of cosmological models. However. this work is based on using independent data points whereas most real world data is correlated in some way. This is normally contained in some covariance matrix. In Markov chain Monte Carlo analyses, this covariance matrix would feature in the log-likelihood of the sampler. Our main aim in the current work is to extend the reconstruction approach of the Hubble diagram to include covariance information. Together with the reconstruction of higher derivatives of the Hubble parameter this means that more complex reconstruction programmes of cosmological models can be considered.

\begin{figure}[h]
	\begin{center}
		
		\begin{tikzpicture}
		
		\foreach \i in {1,...,\inputnum}
		{
			\node[circle, 
			minimum size = 7mm,
			fill=green!50] (Input-\i) at (0,-\i) {};
		}

		\foreach \i in {1,2,3}
		{
			\node[circle, 
			minimum size = 7mm,
			fill=blue!40,
			yshift=(\hiddennum-\inputnum)*5 mm
			] (Hidden-\i) at (2.5,-\i) {};
		}
		
		\node(dots) at (2.5,-1.9){\vdots};
		
		\node[circle, 
		minimum size = 7mm,
		fill=blue!40,
		yshift=(4-\inputnum)*5 mm
		] (Hidden-5) at (2.5,-4) {};

		\foreach \i in {1,...,\hiddennum}
		{
			\node[circle, 
			minimum size = 7mm,
			fill=blue!40,
			yshift=(\hiddennum-\inputnum)*5 mm
			] (Hidden-\i) at (2.5,-\i) {};
		}
		
\foreach \l [count=\i] in {1,2,3,k}
  \node at (Hidden-\i) {$\mathfrak{n}_\l$};
		
		\foreach \i in {1,...,\outputnum}
		{
			\node[circle, 
			minimum size = 7mm,
			fill=red!50,
			yshift=(\outputnum-\inputnum)*5 mm
			] (Output-\i) at (5,-\i) {};
		}
		
		\foreach \i in {1,...,\inputnum}
		{
			\foreach \j in {1,...,\hiddennum}
			{
				\draw[->, shorten >=1pt, color=black!50] (Input-\i) -- (Hidden-\j);	
			}
		}
		
		\foreach \i in {1,...,\hiddennum}
		{
			\foreach \j in {1,...,\outputnum}
			{
				\draw[->, shorten >=1pt,  color=black!50] (Hidden-\i) -- (Output-\j);
			}
		}
		
		\foreach \i in {1,...,\inputnum}
		{            
			\draw[<-, shorten >=1pt] (Input-\i) -- ++(-1,0)
			node[left]{$z$};
		}

		\draw[->, shorten >=1pt] (Output-1) -- ++(1,0)
		node[right]{$\Upsilon(z)$};
		
		\draw[->, shorten >=1pt] (Output-2) -- ++(1,0)
		node[right]{$\sigma_\Upsilon^{}(z)$};
		
		\end{tikzpicture}
		\vskip 0.2cm
		Input Layer~~~~~~~~~~~Hidden Layer~~~~~~~Output Layer 
		
	\end{center}
    \caption{The general structure of the adopted ANN, where the input is the redshift of a cosmological parameter $\Upsilon(z)$, and the outputs are the corresponding value and error of $\Upsilon(z)$.} 
\label{fig:ANN_structure}
\end{figure}
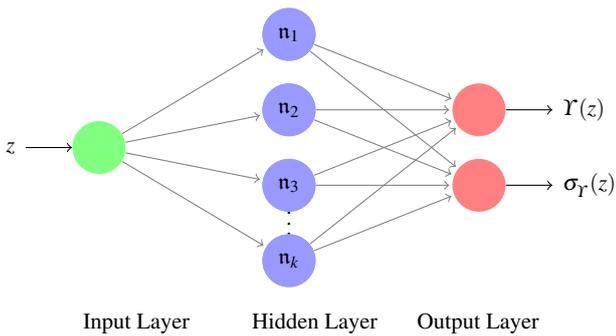


To do this, consider the mechanics of ANN systems in which an input layer is connected to an output layer through a series of hidden internal layers where the majority of neurons are located. These neurons each feature hyperparameters which are set by training with the aim of having new inputs produce outcomes that real observations would. In our setup, the input signal simply consists of a redshift value while the output layer gives the mean Hubble parameter at that redshift together with the uncertainty at that point. This system is depicted in Fig.~\ref{fig:ANN_structure} for a generalized scenario where each redshift value $z$ results in a generic cosmological parameter $\Upsilon(z)$ together with its corresponding uncertainty $\sigma_\Upsilon^{}(z)$.

The ANN architecture is composed of each neuron possessing an \textit{activation function} which calibrates the impact each neuron will have on the output for a particular input signal. Each neuron depends on hyperparameters (weights and biases) which during the training of the ANN take an optimal value. The layers are then structured as the input and output connections between each neuron. In this way, a signal traverses the whole network to produce an output signal in a structured way. In this work, we consider the Exponential Linear Unit (ELU) \cite{2015arXiv151107289C} as the activation function, specified by
\begin{equation}
    f(x) = 
  \begin{cases} 
   {x} & \text{if } x>0 \\
   {\alpha(e^x-1)} & \text{if } x \leq 0
  \end{cases}\,,
\end{equation}
where $\alpha$ is a positive hyperparameter that scales the value to which negative inputs are calibrated to, while positive inputs continue to traverse the network. Thus, complexity in the data would be incorporated through differently optimized hyperparameter values.

The hyperparameter values are set in the training process where real data is inputted through the system and hyperparameter values are optimized against real-world data. This is characterized by a \textit{loss function} which measures the difference between predicted and ground truth values in $\Upsilon$. By minimizing the loss function, the ANN hyperparameters are optimized for particular data sets. An example of this process is the gradient descent combined with the back-propagation algorithm, while Adam's algorithm \cite{2014arXiv1412.6980K} represents a slightly better version of this optimization algorithm. The L1 loss function is the simplest and most direct way of assessing the difference between the predicted and observed values of some parameter, where the absolute difference between observed and predicted values of the Hubble parameter at the observation redshift points are each summed, that is
\begin{equation}
    {\rm L1} = \sum_i |H_{\rm obs}(z_i) - H_{\rm pred}(z_i)|\,,
\end{equation}
where $H_{\rm obs}(z)$ and $H_{\rm pred}(z)$ are observed and ANN predicted values of the Hubble parameter at observation redshifts $z$. This is akin to the MCMC log-likelihood for independent data sets (less the uncertainties). Other loss functions exist but they do not generally incorporate more complexity in the data. In this work, we consider a native way to incorporate more complexity in the observed data sets by defining a new loss function analogous to the MCMC log-likelihood for correlated data sets. We do this by defining the following $\chi^2$ loss function
\begin{equation}
    {\rm L_{\chi^2}} = \sum_{i,j} \left[H_{\rm obs}(z_i) - H_{\rm pred}(z_i)\right]^\text{T} \mathrm{C}_{ij}^{-1} \left[H_{\rm obs}(z_j) - H_{\rm pred}(z_j)\right] \,,
\end{equation}
where $\mathrm{C}_{ij}$ is the total noise covariance matrix of the data, which includes the statistical noise and systematics. In this way, we will be able to naively use correlated data in our ANN architecture. While the exact details of the training process are contained in Sec.~\ref{sec:ANN_results}, this loss function assures that the ANN will infer Hubble expansion values that reflect both the mean observational values as well as the covariance matrix relationships between these points. To ensure the fidelity of this process, we employ a batch size that is equal to the Pantheon compilation sample size. On the other hand, one could divide this matrix and utilize smaller batch sizes if the whole data set were to be unmanageable larger.

\begin{figure*}[ht]
   \centering
    \includegraphics[width=0.45\textwidth]{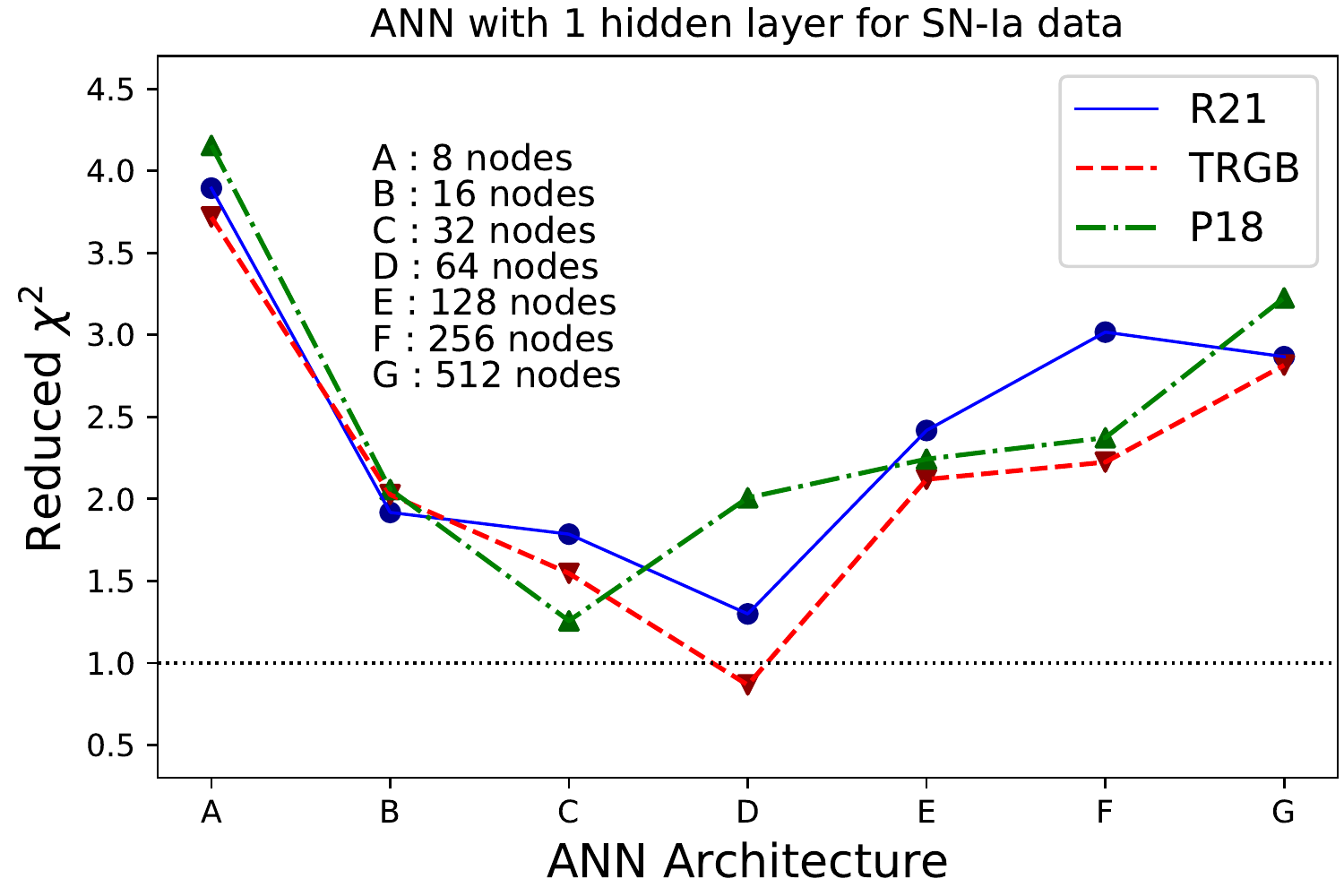}
    \includegraphics[width=0.45\textwidth]{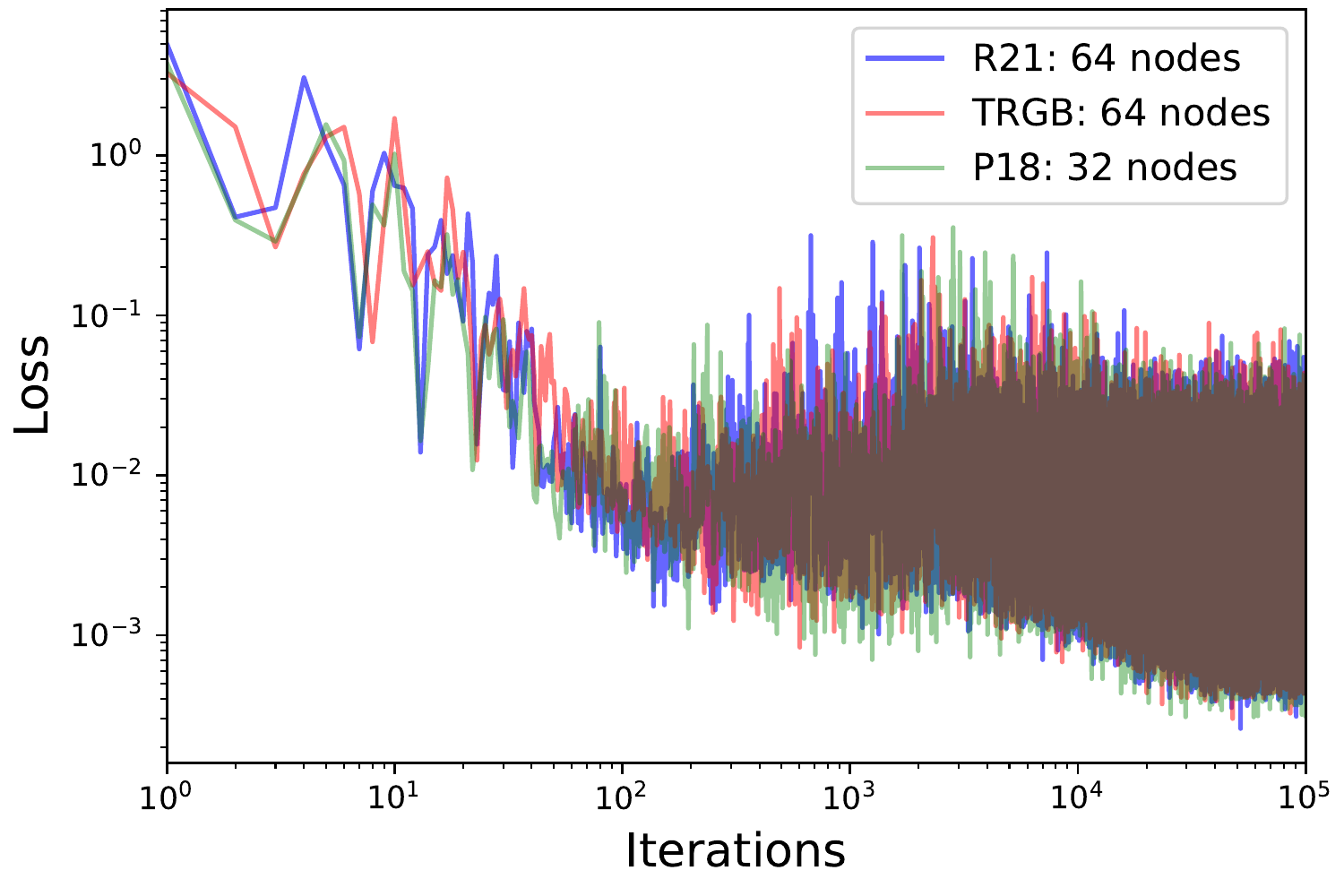} \\ \vskip 0.1cm
    \includegraphics[width=0.45\textwidth]{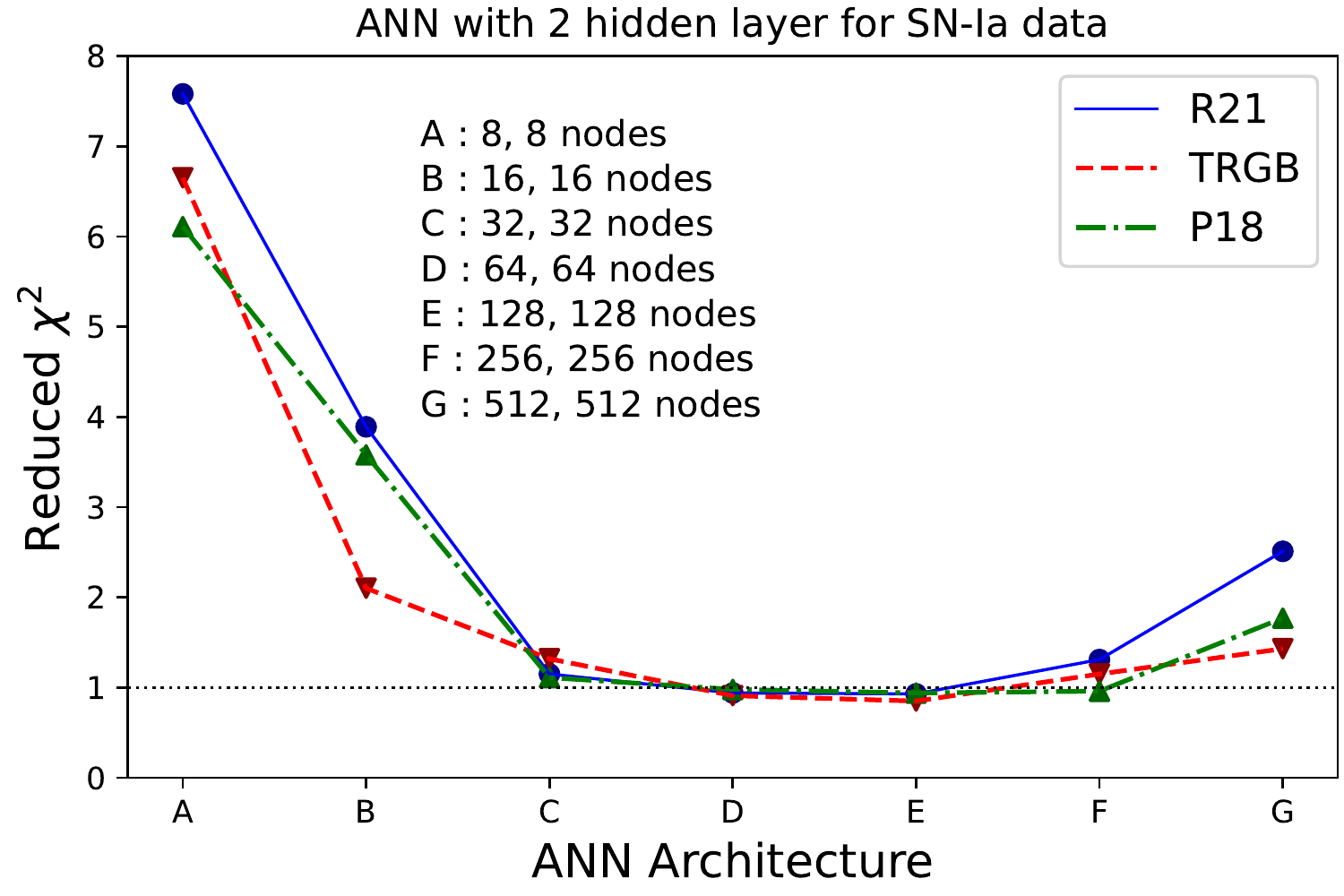}
    \includegraphics[width=0.45\textwidth]{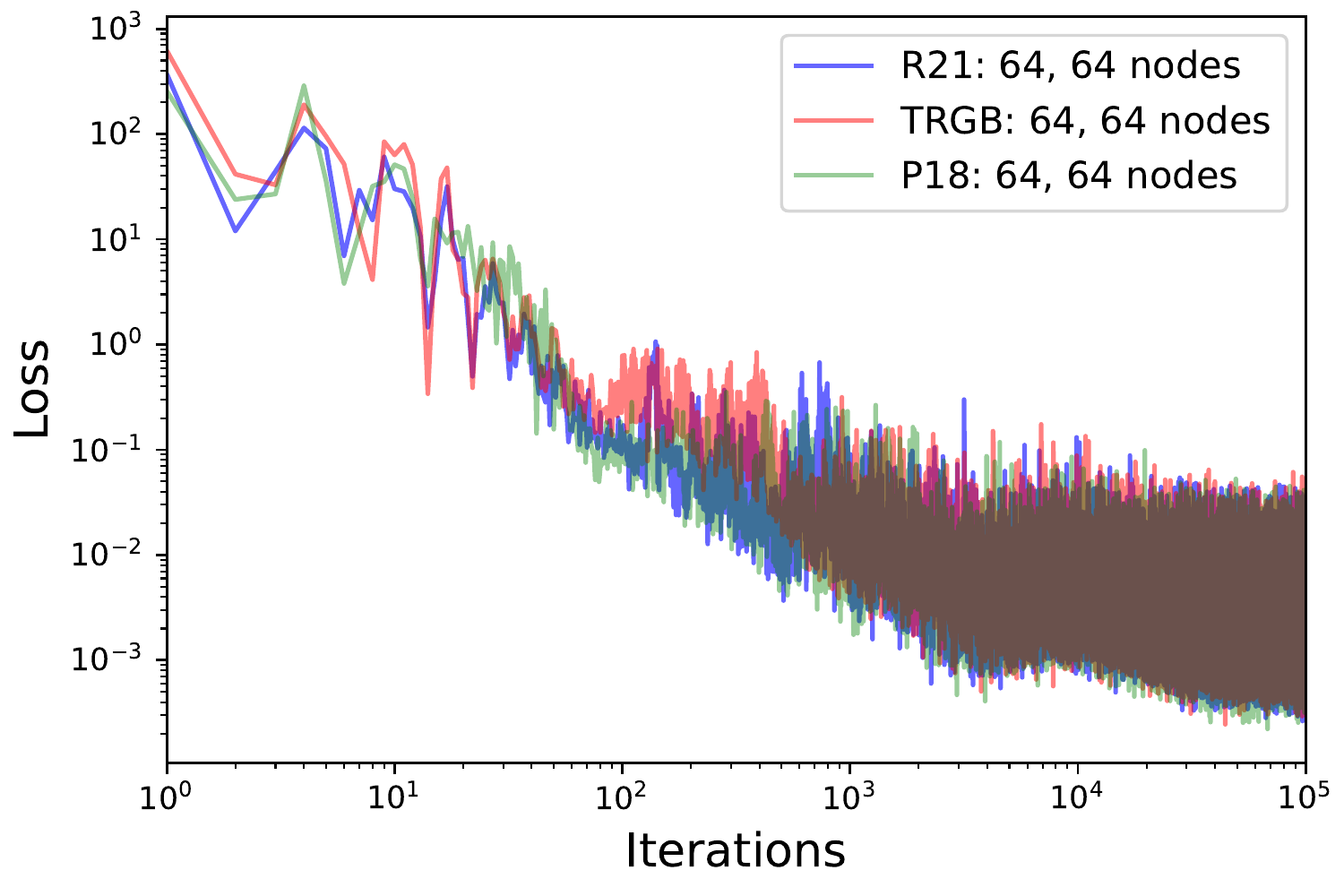}
    \caption{Plots showing the reduced $\chi^2$ (left panel), and the evolution of the $\chi^2$ loss function (right panel), for configuring the optimal neural network architecture using the Pantheon SN-Ia $d_L$ compilation.}
    \label{fig:loss_risk}
\end{figure*}

\begin{table*}[t!]
\begin{center}
\resizebox{\textwidth}{!}{\renewcommand{\arraystretch}{1.05} \setlength{\tabcolsep}{15 pt} \centering
\begin{tabular}{c c c c c c} 
\hline
Layers & Nodes / Neurons & Architecture &  Prior & Average of last 100 Loss & Reduced $\chi^2$ \\
\hline
     &    &   &  R21  &  0.0049 &  3.89   \\
     &  8  &  A &  TRGB  & 0.0045 &  3.72   \\
     &    &   &  P18 &  0.0046 & 4.16 \\
     \cline{4-6}
     &    &   &  R21  & 0.0037 &  1.92   \\
     &  16  &  B &  TRGB  & 0.0035 &  2.03   \\
     &    &   &  P18 &  0.0032 &  2.06 \\
     \cline{4-6}
     &    &   &  R21  &  0.0033 &   1.79  \\
     &  32  &  C &  TRGB  & 0.0032 &  1.55   \\
     &    &   &  P18 &  0.0031 & 1.26 \\
     \cline{4-6}
      &    &   &  R21  &  0.0029 &  1.30   \\
  1    &  64  &  D &  TRGB  & 0.0030 &  0.87   \\
      &    &   &  P18 &  0.0026 & 2.01 \\
      \cline{4-6}
      &    &   &  R21  &  0.0030 &  2.42   \\
      &  128  &  E &  TRGB  & 0.0035 &  2.12   \\
      &    &   &  P18 & 0.0027 &  2.24 \\
      \cline{4-6}
      &    &   &  R21  &  0.0031 &   3.02  \\
      &  256  &  F &  TRGB  & 0.0032 &  2.23   \\
      &    &   &  P18 &  0.0029 &  2.37 \\
      \cline{4-6}
      &    &   &  R21  &  0.0030 &  2.87   \\
      &  512  &  G &  TRGB  & 0.0036 &   2.82  \\
      &    &   &  P18 &  0.0022 &  3.23 \\
     \hline
    &    &   &  R21  &  0.0096 &  7.58   \\
    &  8, 8  &  A &  TRGB  & 0.0101 &  6.65   \\
    &    &   &  P18 &   0.0075 &  6.11 \\
    \cline{4-6}
    &    &   &  R21  &  0.0044 &  3.89   \\
    &  16, 16  &  B &  TRGB  &  0.0056 &  2.11   \\
    &    &   &  P18 &  0.0042  &  3.58 \\
    \cline{4-6}
    &    &   &  R21  &  0.0039 &   1.15  \\
    &  32, 32  &  C &  TRGB  & 0.0041 &   1.32  \\
    &    &   &  P18 &  0.0033 &  1.11 \\
    \cline{4-6}
    &    &   &  R21  & 0.0030 &   \textbf{0.94}  \\
2    &  \textbf{64, 64}  &  D &  TRGB  &  0.0033 &  \textbf{0.91}   \\
    &    &   &  P18 &  0.0024 &  \textbf{0.98} \\
    \cline{4-6}
    &    &   &  R21  &  0.0030 &   0.93  \\
    &  128, 128  &  E &  TRGB  & 0.0031 &  0.85   \\
    &    &   &  P18 &  0.0022 & 0.94 \\
    \cline{4-6}
    &    &   &  R21  &  0.0051 &   1.31  \\
    &  256, 256  &  F &  TRGB  & 0.0044 &   1.15  \\
    &    &   &  P18 &  0.0026 &  0.96 \\
    \cline{4-6}
    &    &   &  R21  &  0.0067 &   2.51  \\
    &  512, 512  &  G &  TRGB  & 0.0063 &   1.43  \\
    &    &   &  P18 & 0.0059 &  1.77 \\
    \hline
\end{tabular}
}
\end{center}
\caption{{\small Reduced $\chi^2$ obtained with different neural network architecture to determine the optimal configuration for the Pantheon SN-Ia $d_L$ data. The best neural network architecture is highlighted in bold.}}
\label{tab:chi2}
\end{table*}

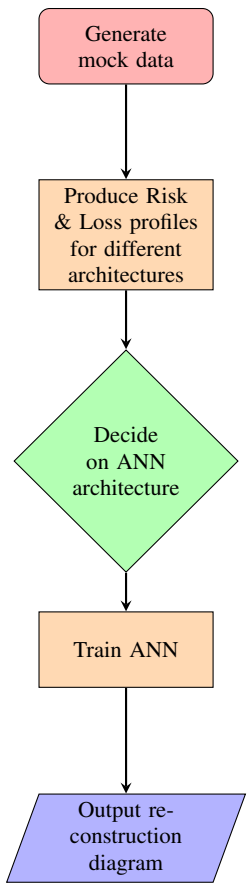
\begin{figure}[ht]
\centering
\begin{tikzpicture}[node distance=2.5cm]
    \node (pro1) [startstop] {Generate mock data};
    \node (pro2) [process, below of=pro1] {Produce Risk \& Loss profiles for different architectures};
    \node (dec1) [decision, below of=pro2] {Decide on ANN architecture};
    \node (pro3) [process, below of=dec1] {Train ANN};
    \node (io1) [io, below of=pro3] {Output reconstruction diagram};
    \draw [arrow] (pro1) -- (pro2);
    \draw [arrow] (pro2) -- (dec1);
    \draw [arrow] (dec1) -- (pro3);
    \draw [arrow] (pro3) -- (io1);
\end{tikzpicture}
\caption{Flow of ANN architecture design and reconstruction process.} \label{fig:flow_chart}
\end{figure}

In order to configure and train our network, we undertake the following steps:
\begin{enumerate}
    \item Designing the neural network: After sorting the observational data sets from low to high redshifts, we use simple ANN, with one input layer (to feed the training redshifts) and one output layer (to predict the reconstructed function). We take into account network models with 1 and 2 hidden layers. The dropout rate is set to 0.2 to prevent it from over-fitting. The number of neurons in the hidden layers is chosen as $2^n$ where $2 \leq n \leq 13$. So the ANN architectures are $1,~ 2^n,~ 1$ for ANN with 1 hidden layer and $1,~ 2^n,~ 2^n,~ 1$ for those with two hidden layers. 
    \item Determining the optimal network configuration: The hyperparameters (weights and biases) of the network are initialized with fixed values. All the ANN configurations are trained after $10^5$ iterations, to ensure that the loss function no longer decreases. We set the initial learning rate to 0.01 which goes on decreasing with the number of iterations and compute the averaged loss of the last 100 iterations. The predictions are made at the training redshifts and evaluate reduced $\chi^2$ for all the architectures considered. The ANN architecture with the least averaged loss of the last 100 iterations, and reduced $\chi^2$ just less than 1 is chosen as the optimal configuration. The optimal network architecture for Pantheon $d_L$ compilation is found to be 1, 64, 64, 1 (see Fig. \ref{fig:loss_risk} and Table \ref{tab:chi2}). On proceeding in a similar fashion, we get 1, 1024, 1 as the optimal network structure for the Hubble $H(z)$ data.
    \item Monte Carlo approach for final predictions: This optimal network architecture is now iterated over 500 times, for random initialization of hyperparameters along with the dropout effect. Thus, we get 500 samples of the reconstructed functions at the corresponding test redshifts, from which we compute the mean function and the respective uncertainties. 
    \item Derivative predictions: With the 500 realizations of the predicted functions, we compute numerical derivatives as, $ f'(z_i) \simeq \frac{f(z_{i+1}) - f(z_{i-1})}{z_{i+1} - z_{i-1}} $.    From the reconstructed $f'(z)$ samples, we obtain the mean values of reconstructed $f'(z)$ along with the associated confidence levels using another MC routine \cite{Mukherjee:2022yyq}.   
    \item Batch size: For determining the optimal network configuration, we employ a batch size that is equal to the data size. During the final predictions, the batch size adopted for the Pantheon compilation is 40 (equal to the size of the binned Pantheon data), and half the number of available measurements for the Hubble data.
    \end{enumerate}
These are also illustrated in Fig.~\ref{fig:flow_chart} where the different processes in the construction, training and eventual reconstruction procedures are connected together.

\subsection{Data sets}

We now employ ANNs to reconstruct the Hubble diagram, considering three sources of data. These include the cosmic chronometers (CC) and baryonic acoustic oscillation (BAO) measurements of the Hubble parameter, as well as the type Ia supernovae (SN) apparent magnitude data. Furthermore, keeping in mind the rising $H_0$ tension, we consider the most precise Cepheid calibration result of $H_0 =  73.3 \pm 1.04$ km Mpc$^{-1}$ s$^{-1}$ \cite{Riess:2021jrx} by the SH0ES team (hereafter referred to as R21), recently inferred $H_0 = 69.7 \pm 1.9$ km Mpc$^{-1}$ s$^{-1}$ \cite{Freedman:2021ahq} via the Tip of the Red Giant Branch (TRGB) calibration technique (hereafter referred to as TRGB) and the most precise early-time determination of $H_0 = 67.4 \pm 0.5$ km Mpc$^{-1}$ s$^{-1}$ \cite{Aghanim:2018eyx} inferred from the Cosmic Microwave Background (CMB) sky by the Planck 2018 survey (hereafter referred to as P18).  In our analysis, we assume Gaussian prior distributions with the mean and variances corresponding to the central and 1$\sigma$ reported values of each prior above.

\begin{figure}[ht]
\centering
\includegraphics[width=0.45\textwidth]{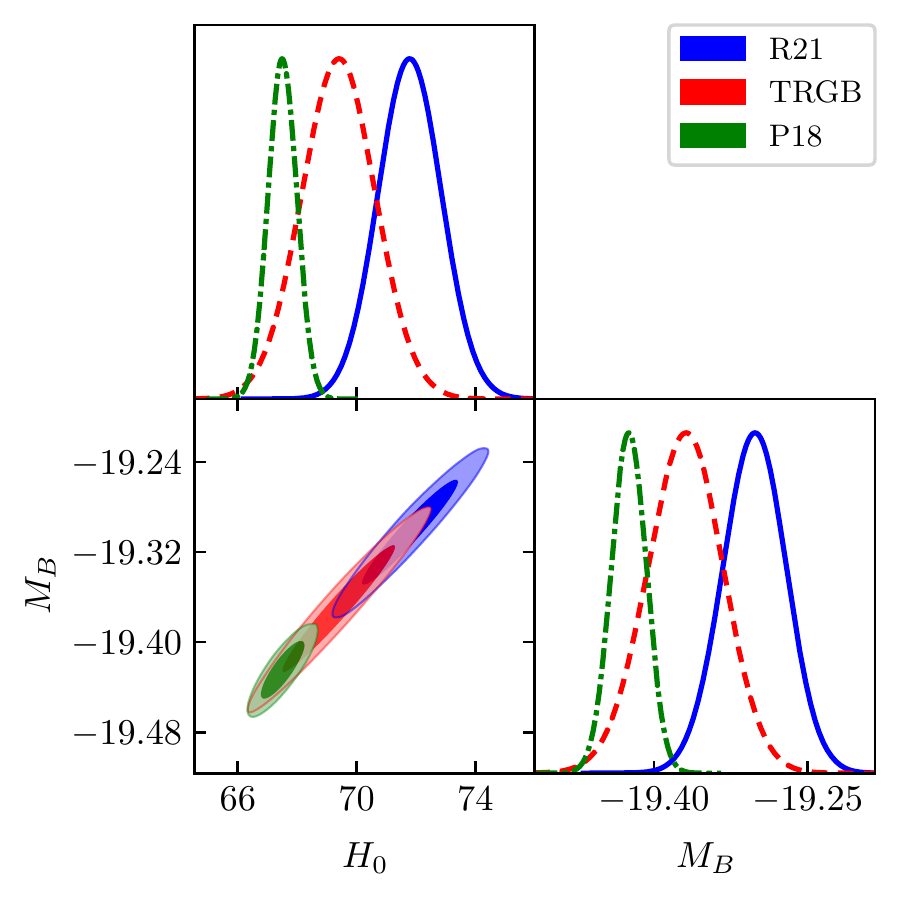}  
\caption{Marginalized posteriors for the calibrated values of supernovae apparent magnitude $M_B$ in the Pantheon compilation considering the R21, TRGB, and P18 $H_0$ priors (in units of km Mpc$^{-1}$ s$^{-1}$), respectively. The constraints obtained are $M_B$ = $-19.302 \pm 0.031$, $-19.369 \pm 0.037$ and $-19.425\pm 0.017$ corresponding to the R21, TRGB and P18 $H_0$ priors.} \label{fig:MB_calib}
\end{figure}

For the SN data, we take into account the full Pantheon \cite{Pan-STARRS1:2017jku} compilation consisting of 1048 supernovae. We attempt to reconstruct the comoving distances from the Pantheon compilation. To begin with, we convert the apparent magnitudes $m(z)$ from the full supernova sample to the respective luminosity distances (in units of Mpc), as  
\begin{equation}
    d_L(z) = {10^{\frac{1}{5} \left[ m(z) - M_B -25 \right]}},
\end{equation} where $M_B$ is the absolute magnitude of supernovae. We obtain the marginalized constraints on $M_B$ assuming vanilla $\Lambda$CDM, considering a uniform prior $M_B \in [-35, -5]$ via a Markov Chain Monte Carlo (MCMC) analysis using {\texttt{emcee}}\footnote{\url{https://github.com/dfm/emcee}} \cite{Foreman-Mackey:2012any} python library. The calibrated constraints obtained are $M_B$ = $-19.302 \pm 0.031$, $-19.369 \pm 0.037$ and $-19.425\pm 0.017$ corresponding to the R21, TRGB and P18 $H_0$ priors, respectively, are shown in Fig. \ref{fig:MB_calib} using {\texttt{GetDist}}\footnote{\url{https://github.com/cmbant/getdist}} \cite{Lewis:2019xzd}. 

Again, we make use of the latest 32 CC Hubble parameter measurements \cite{Stern:2009ep,Moresco:2012jh, Moresco:2016mzx, Borghi:2021rft, Ratsimbazafy:2017vga, Moresco:2015cya, Zhang:2012mp}, covering the redshift range up to $z \sim 2$. These data do not assume any particular cosmological model but depend on the differential ages technique between galaxies, where we consider the full covariance matrix including the systematic and calibration errors \cite{Moresco:2020fbm}. We also take into account the BAO Hubble distance $\frac{d_H(z)}{r_d}$ measurements \cite{BOSS:2016wmc,Bautista:2020ahg,Gil-Marin:2020bct,Neveux:2020voa,Hou:2020rse,deSainteAgathe:2019voe,Blomqvist:2019rah}  from different galaxy surveys like Sloan Digital Sky Survey (SDSS), the Baryon Oscillation Spectroscopic Survey (BOSS) and the extended Baryon Oscillation Spectroscopic Survey (eBOSS), such that
\begin{equation}
    H(z) = c/{d_H(z)}. 
\end{equation} 
Note that, the BAO $H(z)$ data assumes a fiducial value for the radius of the comoving sound horizon $r_d$.  To investigate the effect of the sound horizon scale on the reconstruction when using the BAO data, we consider the following constraint on $r_d h = 102.56 \pm 1.87$ obtained by Camarena \& Marra \cite{Camarena:2019rmj}, keeping in mind the degeneracy between $H_0$ and $r_d$.

\section{Neural Network Reconstruction} \label{sec:ANN_results}

After preparation of the $d_L$ data, we train a network model to learn to mimic the complex relationships between $z$, $d_L(z)$ and $\sigma_{d_L}(z)$. With this trained model, any arbitrary number of $d_L(z)$ samples can be reconstructed by feeding a sequence of redshifts to this network model. Before training the network model on real data, we structure the optimal network configuration of our network model, i.e. determining the optimal number of neurons and layers according to Sec. A of \cite{Mukherjee:2022yyq}.

Now, for the given sample of reconstructed $d_L(z)$,  we can arrive at the evolution of the normalized transverse comoving distance, $D$, from the Pantheon sample as
\begin{equation}
    D(z) = \frac{H_0}{c(1+z)} d_L(z)\,. \label{eq:norm_trans_com_dist}
\end{equation}

\begin{figure}[ht]
\centering
\includegraphics[width=0.45\textwidth]{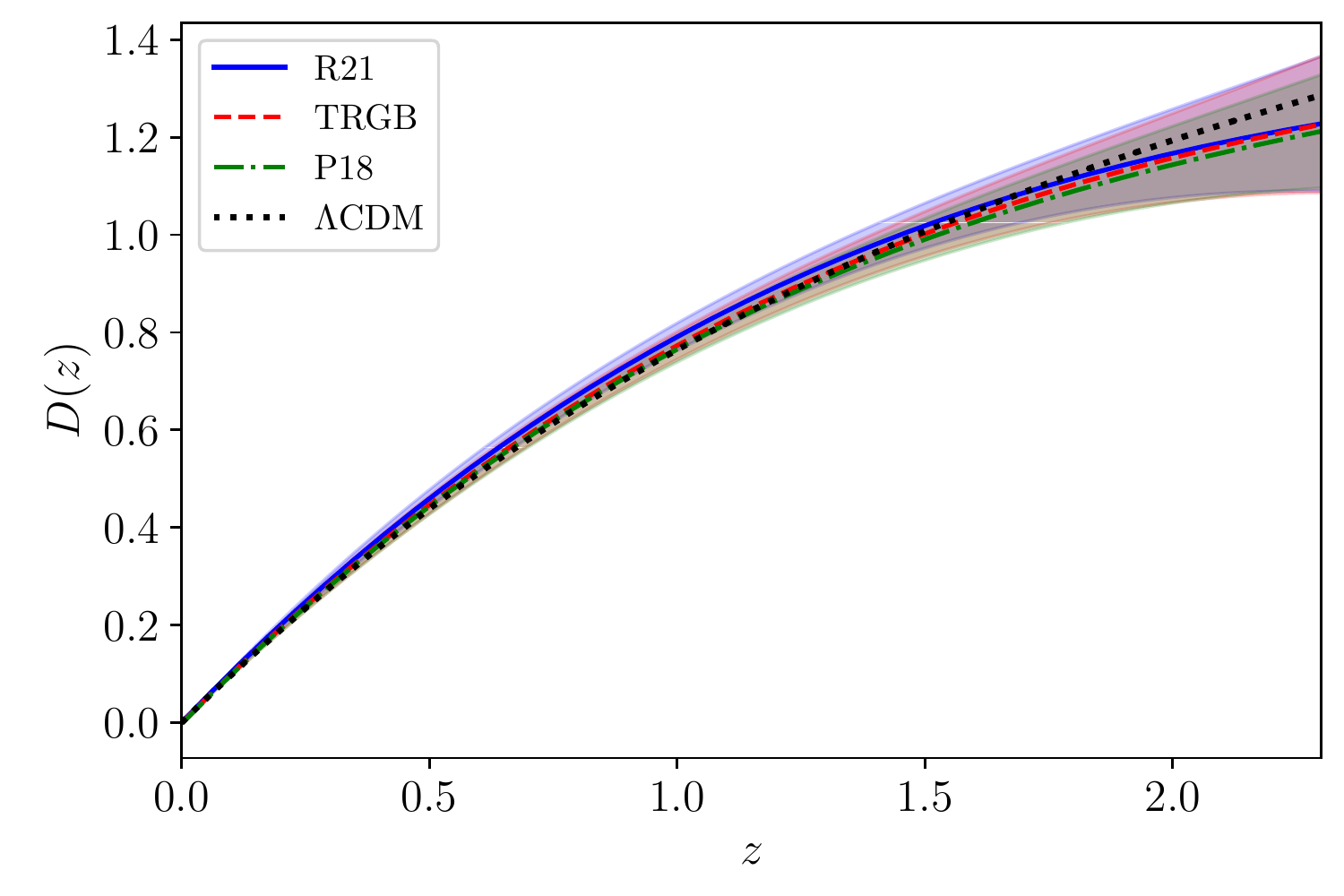}
\includegraphics[width=0.45\textwidth]{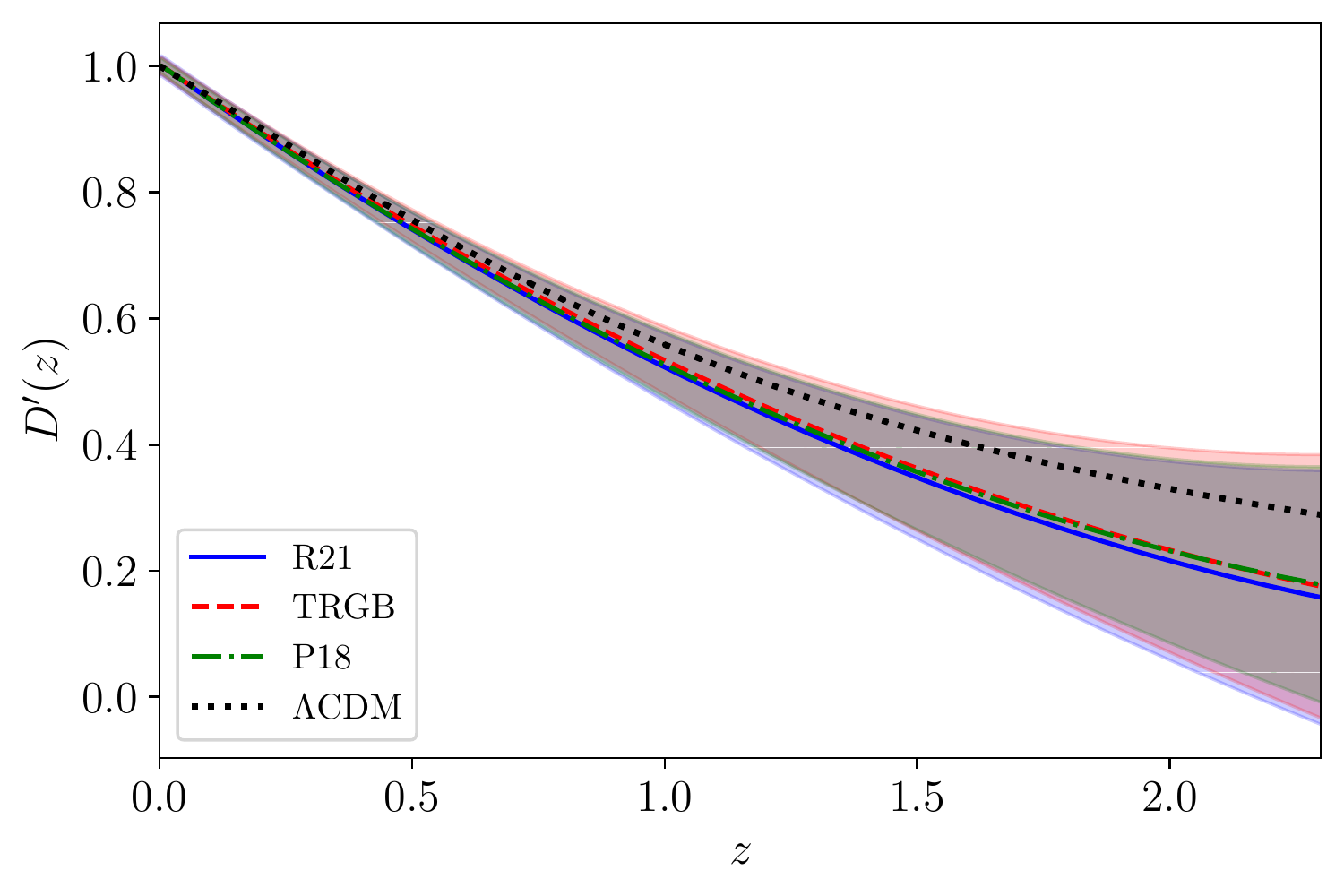}
\caption{Plots for the reconstructed (i) $D(z)$ (left panel), and (ii) $D^\prime(z)$ (right panel), using neural networks from the Pantheon SN data considering R21, TRGB, and P18 $H_0$ priors. }
\label{fig:sn_recon_nn}
\end{figure}

\begin{figure}[ht]
\centering
\includegraphics[width=0.45\textwidth]{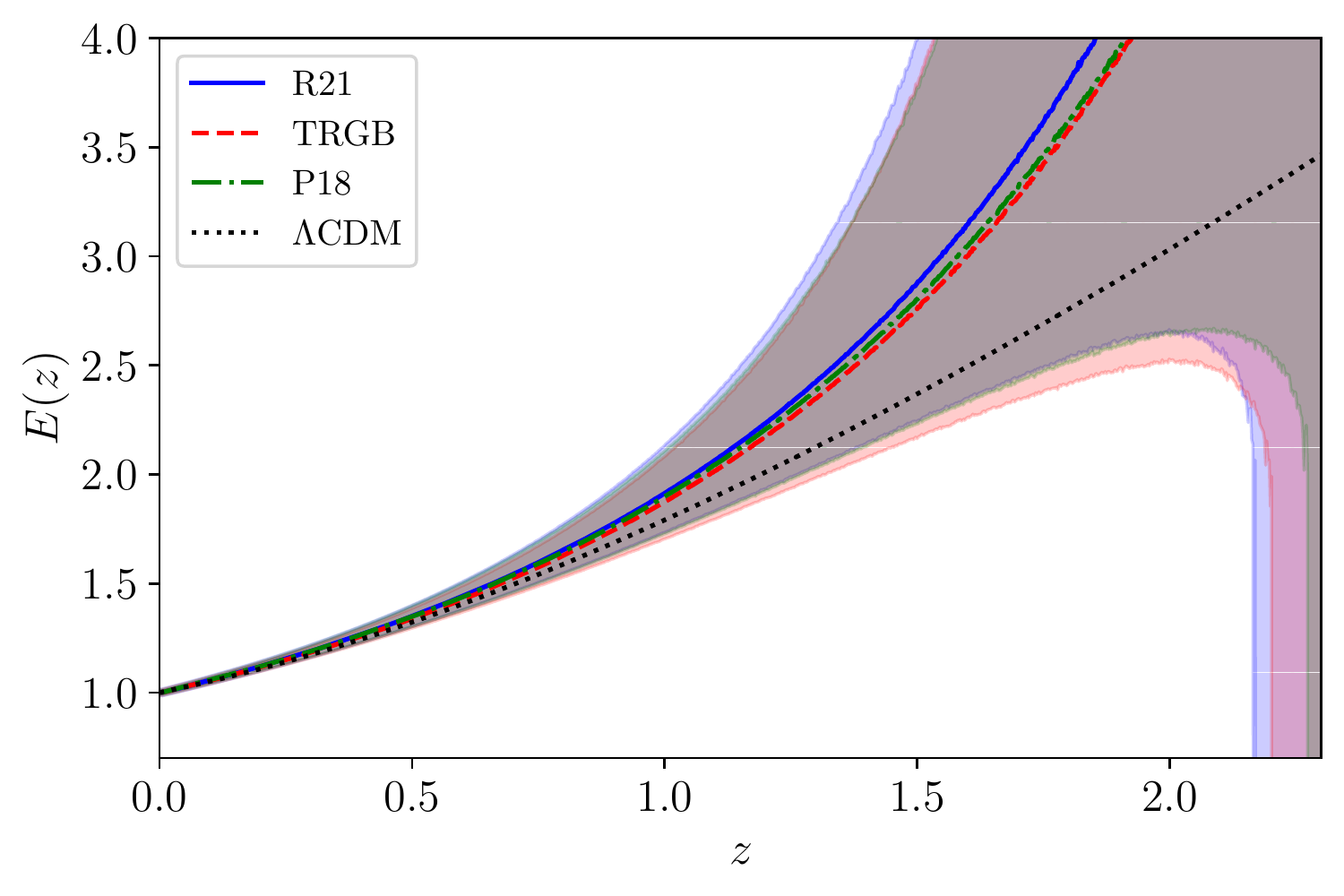}
\includegraphics[width=0.45\textwidth]{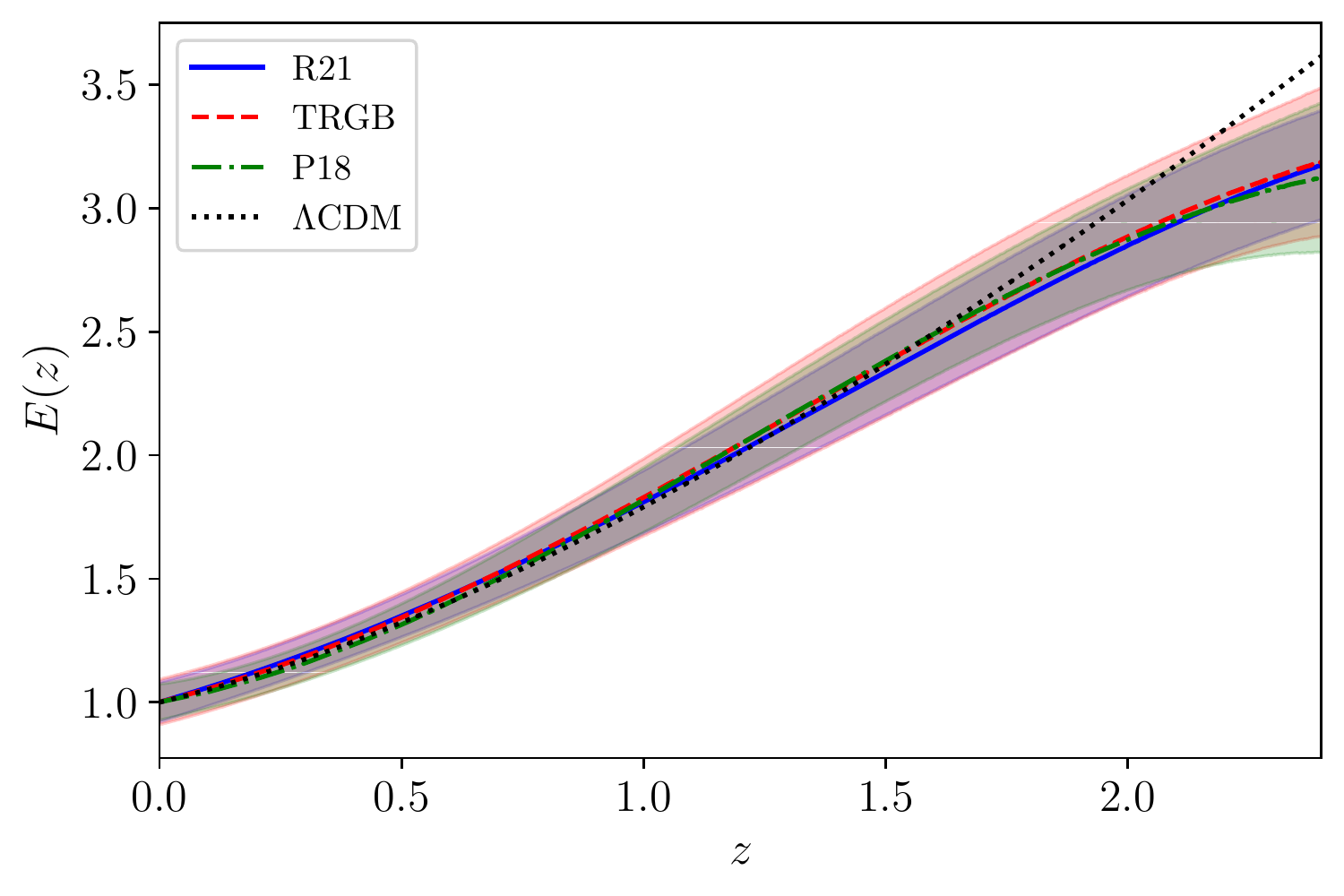}
\caption{Plots for the reconstructed reduced Hubble parameter $E(z)$ from the (i) Pantheon SN compilation (left panel) and (ii) combined CC+BAO Hubble data set (right panel), using neural networks considering R21, TRGB, and P18 $H_0$ priors.}
\label{fig:E_recon_nn}
\end{figure}

The plot for the reconstructed $D$ is shown in the left panel of Fig. \ref{fig:sn_recon_nn}. In this setting, the reconstruction is produced by feeding a number of redshift points into the ANN so that values of $D$ and its associated uncertainty can be obtained. The observational covariance information will have been imprinted on the ANNs through the training process due to the form of the loss function, while the reconstructed diagram will simply be composed of mean values and uncertainties at specific redshift points. We also undertake the simultaneous reconstruction of $D^\prime(z)$, the first order derivative of $D(z)$, where this prime denotes derivative with respect to the redshift $z$, via an MC routine on multiple $d_L(z)$ realizations, such that $D^\prime(z) = \frac{H_0}{c(1+z)} ~d_L^\prime(z)$. This compounding effect of MC with ANNs is undertaken following the methodology described in Ref. \cite{Mukherjee:2022yyq}.  The plot for the reconstructed $D^\prime(z)$ is shown in the right panel of Fig. \ref{fig:sn_recon_nn}. Finally, one can plot the evolution of the reduced Hubble parameter $E(z)$ from the supernovae data as, $E(z) = 1/D^\prime(z)$, given in the left panel of Fig. \ref{fig:E_recon_nn}.

For a comparison between the Hubble and supernovae data sets, we next utilize the ANN method to reconstruct the reduced Hubble parameter, 
\begin{equation}
    E(z) = H(z)/H_0\,,\label{eq:reduced_hubb}
\end{equation} 
directly from the combined CC+BAO Hubble data. The uncertainty associated with the reconstructed $E(z)$ is obtained via the Monte Carlo method. Plots for the reconstructed $E(z)$ from the Hubble data are shown in the right panel of Fig. \ref{fig:E_recon_nn}.

\begin{figure}[ht]
\centering
\includegraphics[width=0.45\textwidth]{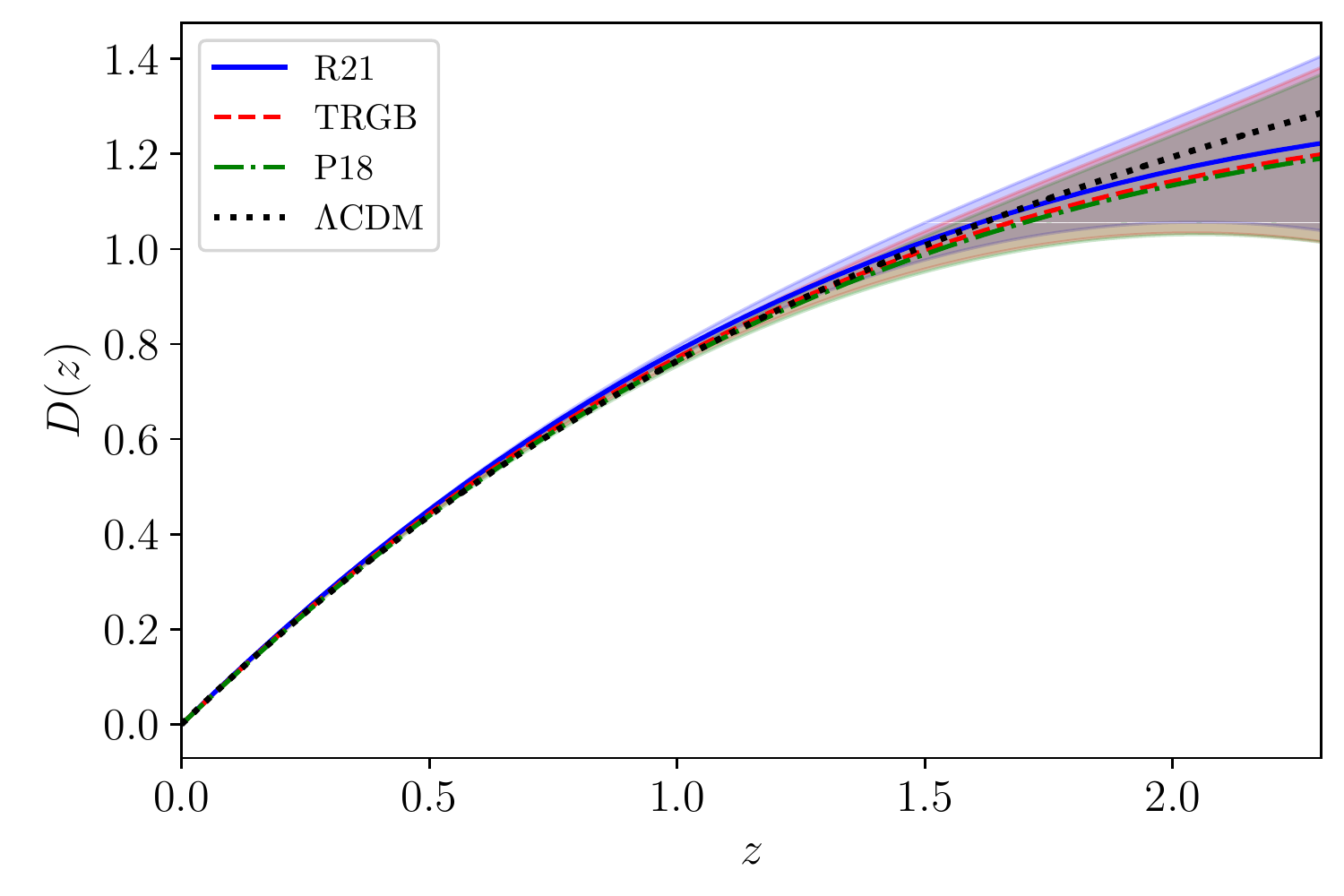}
\includegraphics[width=0.45\textwidth]{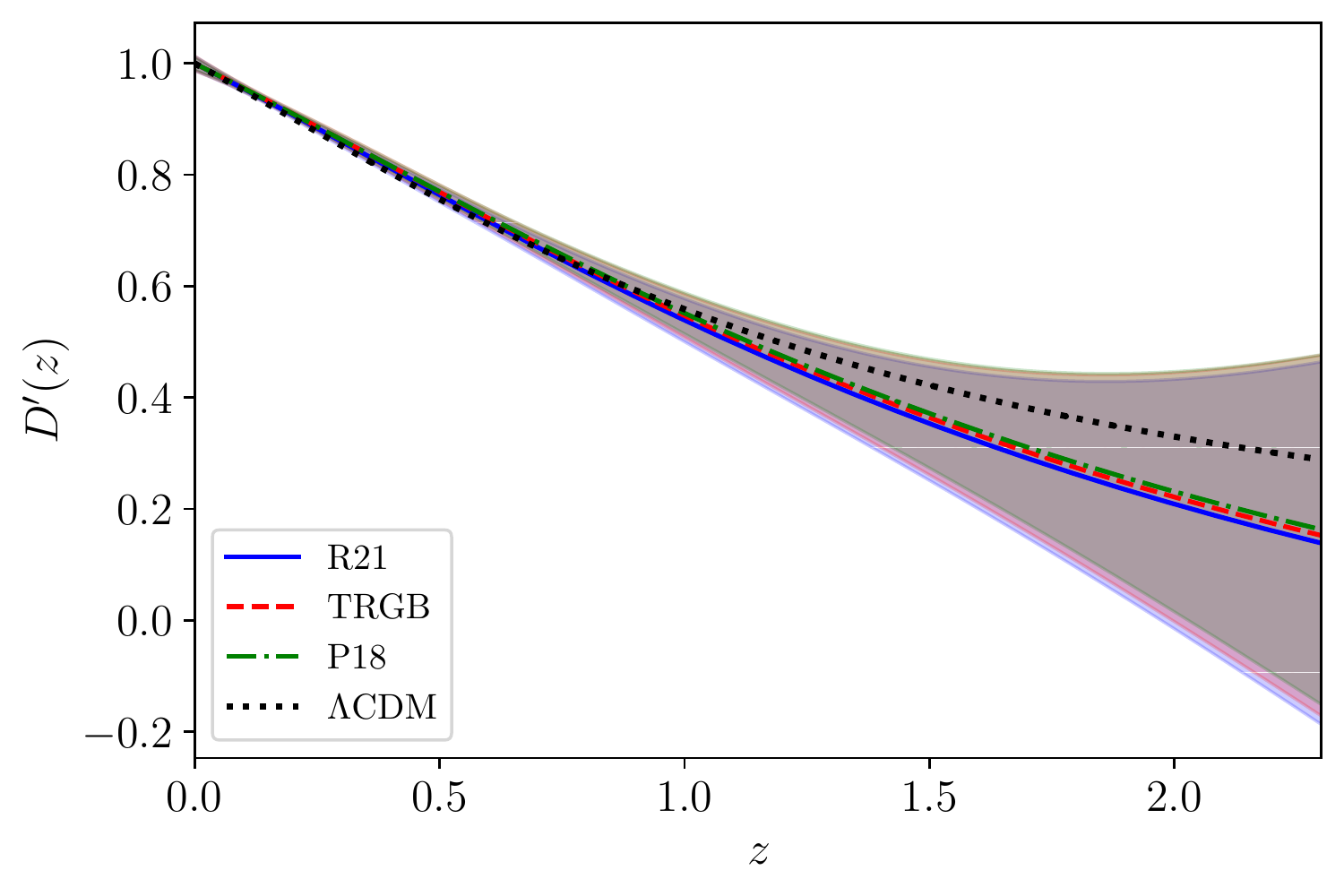}
\caption{Plots for the reconstructed (i) $D(z)$ (left panel), and (ii) $D^\prime(z)$ (right panel), using Gaussian processes from the Pantheon SN data considering R21, TRGB, and P18 $H_0$ priors. }
\label{fig:sn_recon_gp}
\end{figure}

\begin{figure}[ht]
\centering
\includegraphics[width=0.45\textwidth]{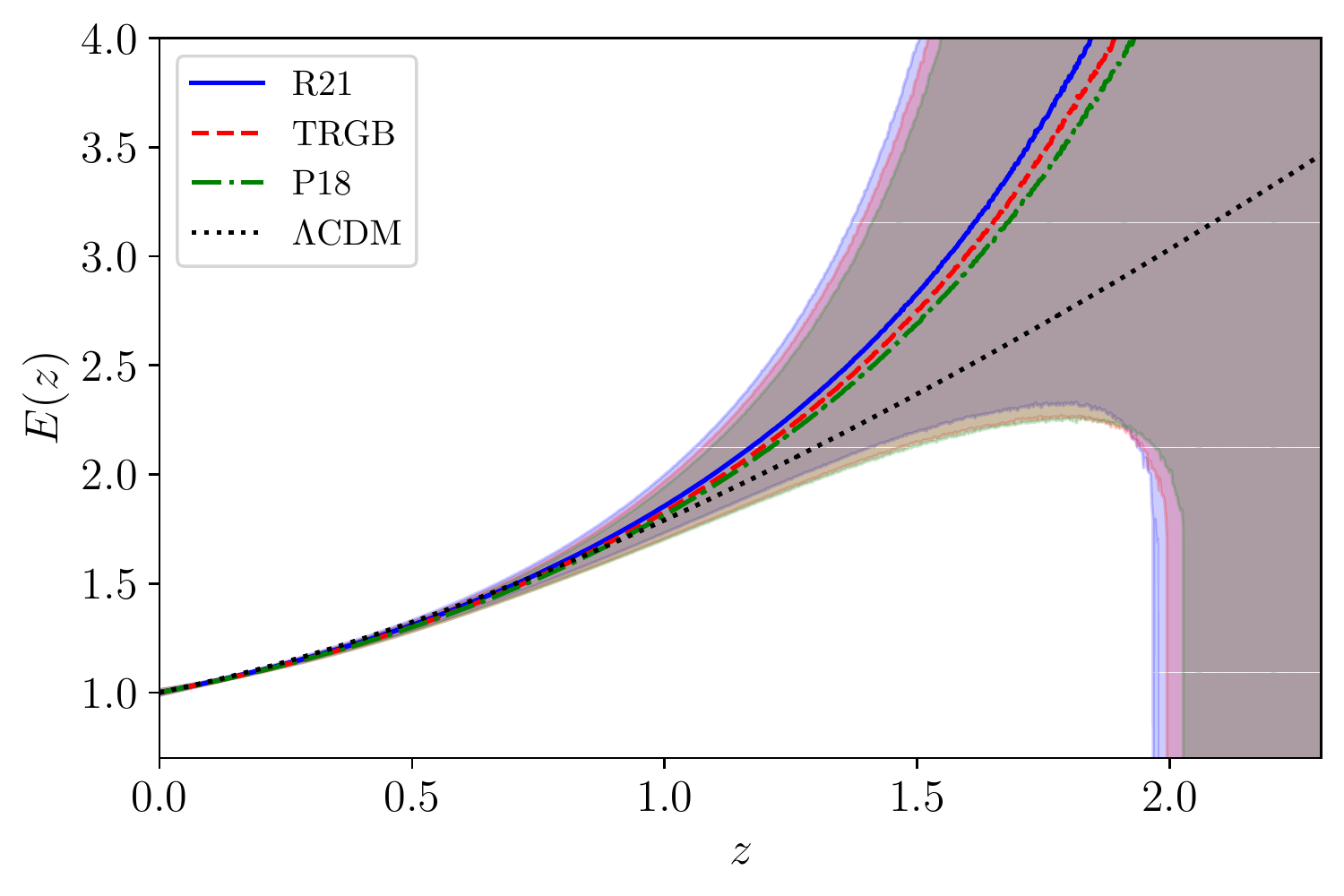}
\includegraphics[width=0.45\textwidth]{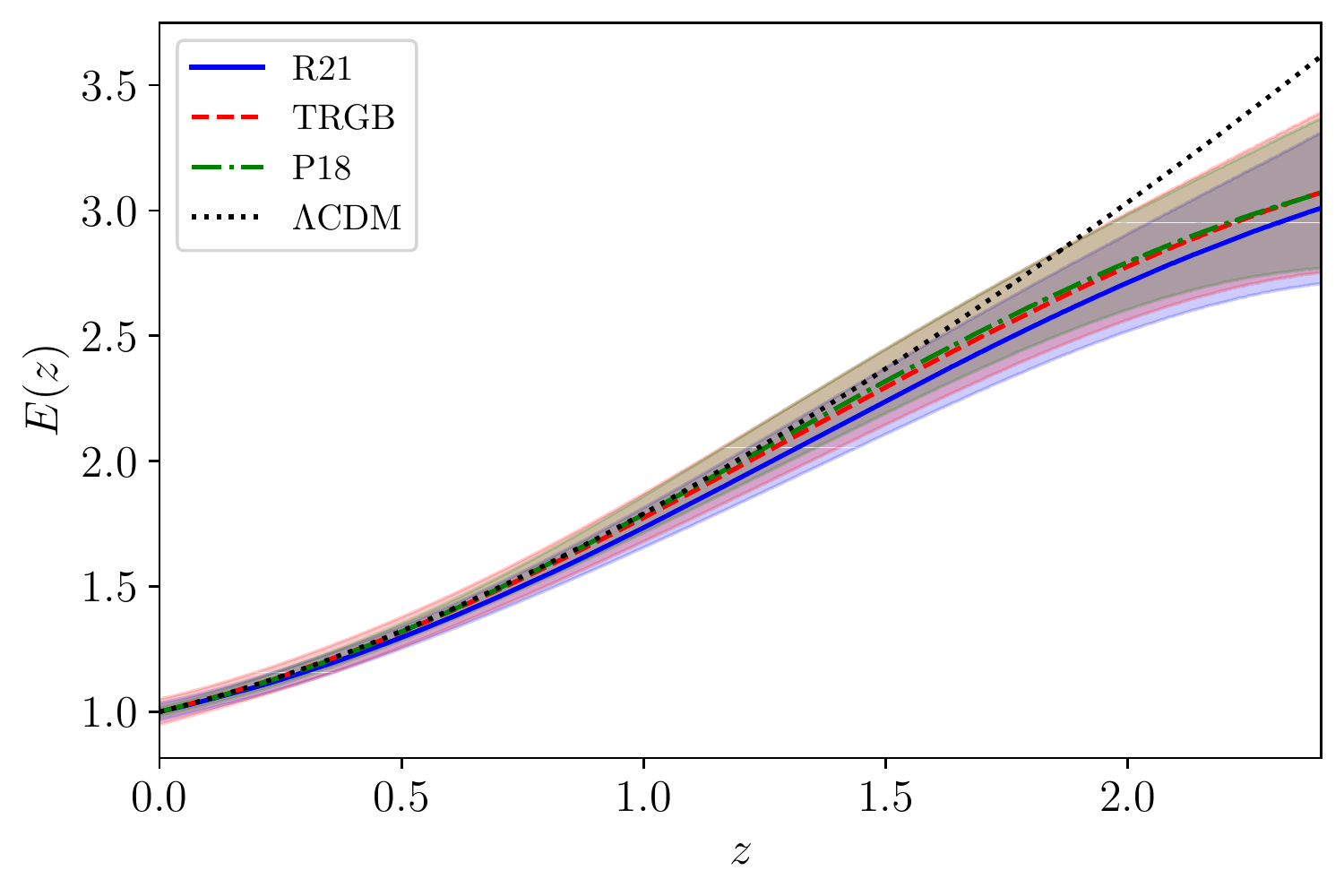}
\caption{Plots for the reconstructed reduced Hubble parameter $E(z)$ from the (i) Pantheon SN compilation (left panel) and (ii) combined CC+BAO Hubble data set (right panel), using Gaussian processes considering R21, TRGB, and P18 $H_0$ priors.}
\label{fig:E_recon_gp}
\end{figure}

\section{Comparison with Gaussian Processes Reconstruction \label{sec:GP}}

In this section, we will discuss the work done in this paper using ANN-based reconstruction techniques, compared to the ones from Gaussian Processes. We recall that the methods by which these two reconstruction strategies function are fundamentally different. While GP requires some constraints on the type of data that it can be applied to, ANNs make vastly fewer assumptions and feature a much higher number of hyperparameters, which are then fit during the training of the neural network. Thus, one would expect an ANN to be much less constrained by the complexity of the data, and to have wider uncertainties. On the other hand, since GP does have some information about the behavior of the data, it can obtain smaller uncertainties.

We start by comparing the normalized transverse comoving distance $D(z)$~\eqref{eq:norm_trans_com_dist} which quantifies the comoving distance for an object of relatively small characteristic length with respect to the Hubble flow. This is an appropriate way in which to interpret the SN data, since it does not require a fully determined cosmological model on which to perform numerical integrals. In our case, we first show the reconstruction for $D(z)$ in Fig.~\ref{fig:sn_recon_nn} where the evolution is shown for a wider range of redshifts with means being shown for the $\Lambda$CDM model, as well as reconstructions for various literature priors. Given our reconstruction approach, we can also show the reconstruction of the redshift derivative of $D(z)$ for the same priors. This can be contrasted with the analogous plot Fig.~\ref{fig:sn_recon_gp} which is the GP reconstruction of the same plots. In both cases, the reconstructions have very low uncertainties for most of the evolution of both $D(z)$ and its first derivative. This happens because there is such a volume of data for the Pantheon sample. Thus, both methods will function quite well in the reconstruction of this particular data set.

The other comparison that provides an important dimension to the performance of GP and ANNs is that of the reduced Hubble parameter described in Eq.~\eqref{eq:reduced_hubb} which is a rescaled Hubble parameter that features a theoretical prior in that $E(0) = 1$. This rescaled Hubble parameter is used for both the Pantheon data set as well as  for Hubble data in the form of CC+BAO. For the ANN reconstruction, the reduced Hubble parameter gives Fig.~\ref{fig:E_recon_nn} in which the reconstruction based on the Pantheon data set shows good behavior for low to medium values of redshift but then becomes numerically unbounded for much larger redshifts, while the same parameter is well behaved for the whole data range in the CC+BAO case. On the other hand, the GP reconstruction, shown in Fig.~\ref{fig:E_recon_gp} has associated uncertainties that increase at slightly lower redshifts for the Pantheon data set case. Also, the CC+BAO reconstruction with is in mild tension with $\Lambda$CDM at comparatively lower redshifts.

GP and ANN both have positive features in reconstructing cosmological data sets. However, ANN shows greater promise in that they rely on less rigid training data and can model more complex structures of data sets.

\section{Null Tests \label{sec:null_tests}}

We now introduce some null tests, namely the $\mathcal{O}m$ diagnostics \cite{Sahni:2008xx,Zunckel:2008ti,Shafieloo:2009hi}, followed by the  $H_0$ diagnostics \cite{Krishnan:2020vaf}, to test the validity of the concordance model of cosmology. 

\subsection{\texorpdfstring{$\mathcal{O}m$}{} diagnostics} \label{sec:om-diagnostics}

The $\mathcal{O}m$ diagnostic \cite{Sahni:2008xx,Zunckel:2008ti,Shafieloo:2009hi} serves as a null test to distinguish the $\Lambda$CDM model from alternative dark energy and modified gravity models, defined as 
\begin{equation} \label{eq:null_test}
    \mathcal{O}m (z) = \frac{E^2(z) - 1}{(1+z)^3 -1}\,,
\end{equation} 
where $E (z) = {H(z)}/{H_0}$ is the reduced Hubble parameter. It works on the principle that different models have different evolutionary trajectories in $z-\mathcal{O}m(z)$ plane. Being a function of $H(z)$ only, which can be directly reconstructed from observational data, it is independent of the cosmic equation of state. Moreover, there is no dependence on any theory of gravity. So, this exercise serves as an alternative route towards understanding the late-time cosmic acceleration in the absence of any convincing physical theory \cite{Clarkson:2007pz,Qi:2016wwb,Qi:2018pej,Bengaly:2020neu}.

For a universe with an underlying expansion history $E(z)$, given by the $\Lambda$CDM model, $\mathcal{O}m(z)$ will essentially be a constant, exactly equal to $\Omega_{m0}$, the matter density parameter at the present epoch. The slope of $\mathcal{O}m(z)$ can differentiate between different dark energy and modified gravity models even if the $\Omega_{m0}$ is not accurately known. Therefore, any possible deviation of $\mathcal{O}m(z)$ from $\Omega_{m0}$ can be used to draw inferences on the dynamics of the universe. For the phenomenological $w$CDM model, where the dark energy component is described by a constant equation of state parameter $w$, a positive slope of the $\mathcal{O}m(z)$ indicates a phantom behaviour of dark energy, whereas a negative slope points towards a quintessence dark energy model.

\begin{figure}[ht]
\centering
\includegraphics[width=0.45\textwidth]{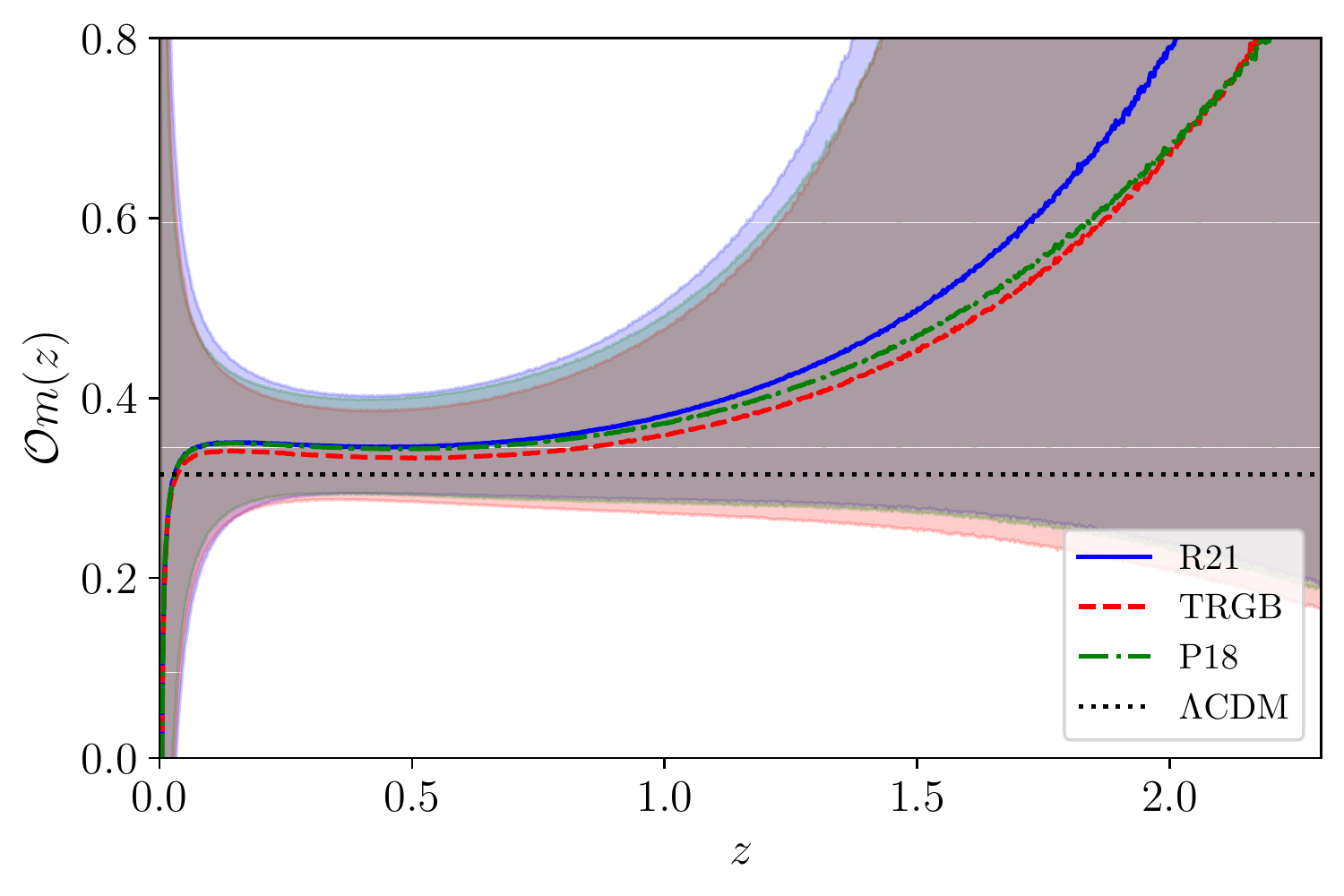}
\includegraphics[width=0.45\textwidth]{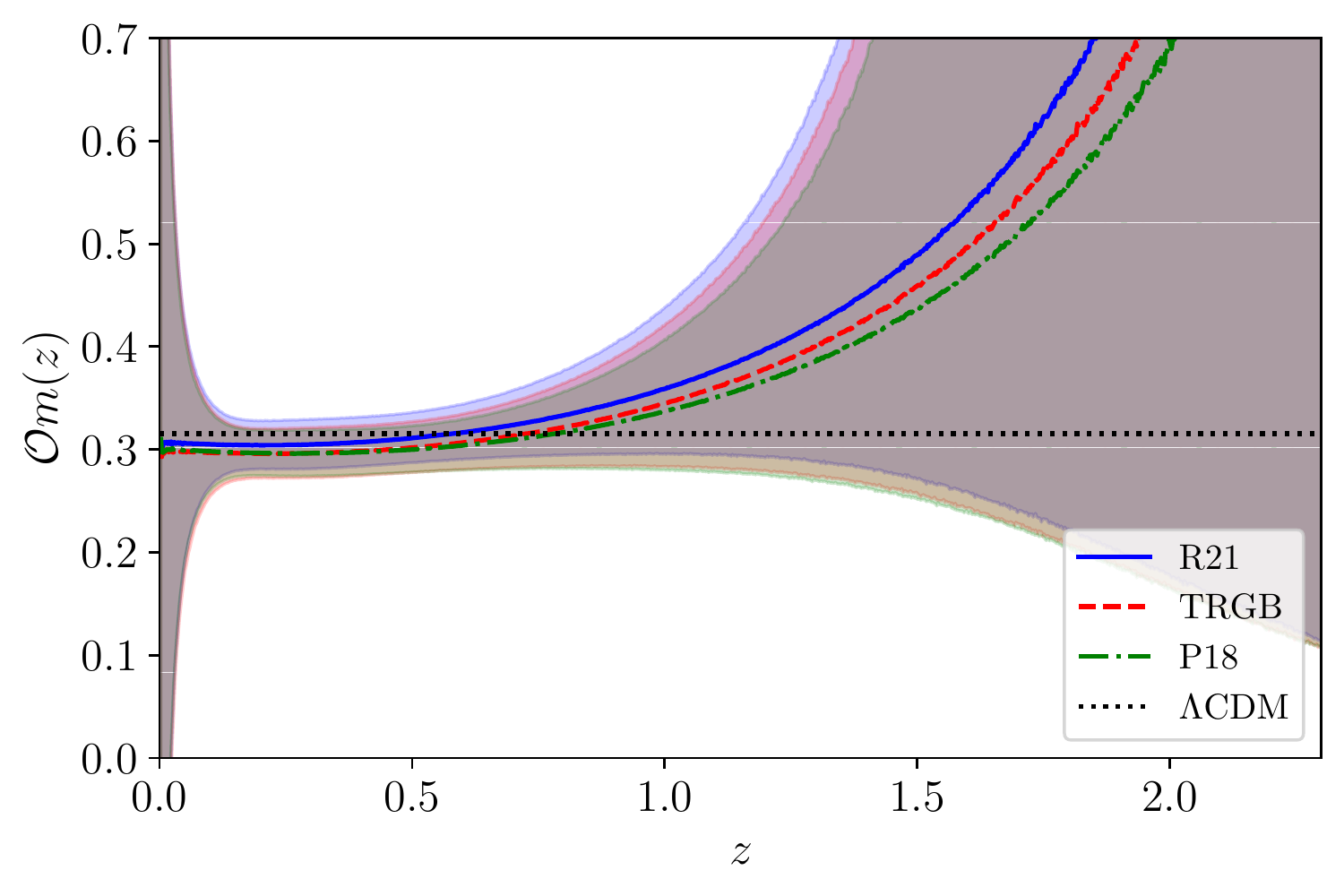}
\caption{Plots for the reconstructed $\mathcal{O}m$ diagnostics using (i) neural networks (left panel) and (ii) gaussian processes (right panel), from the Pantheon SN data considering R21, TRGB, and P18 $H_0$ priors. }
\label{fig:om_diag_sn}
\end{figure}

\begin{figure}[ht]
\centering
\includegraphics[width=0.45\textwidth]{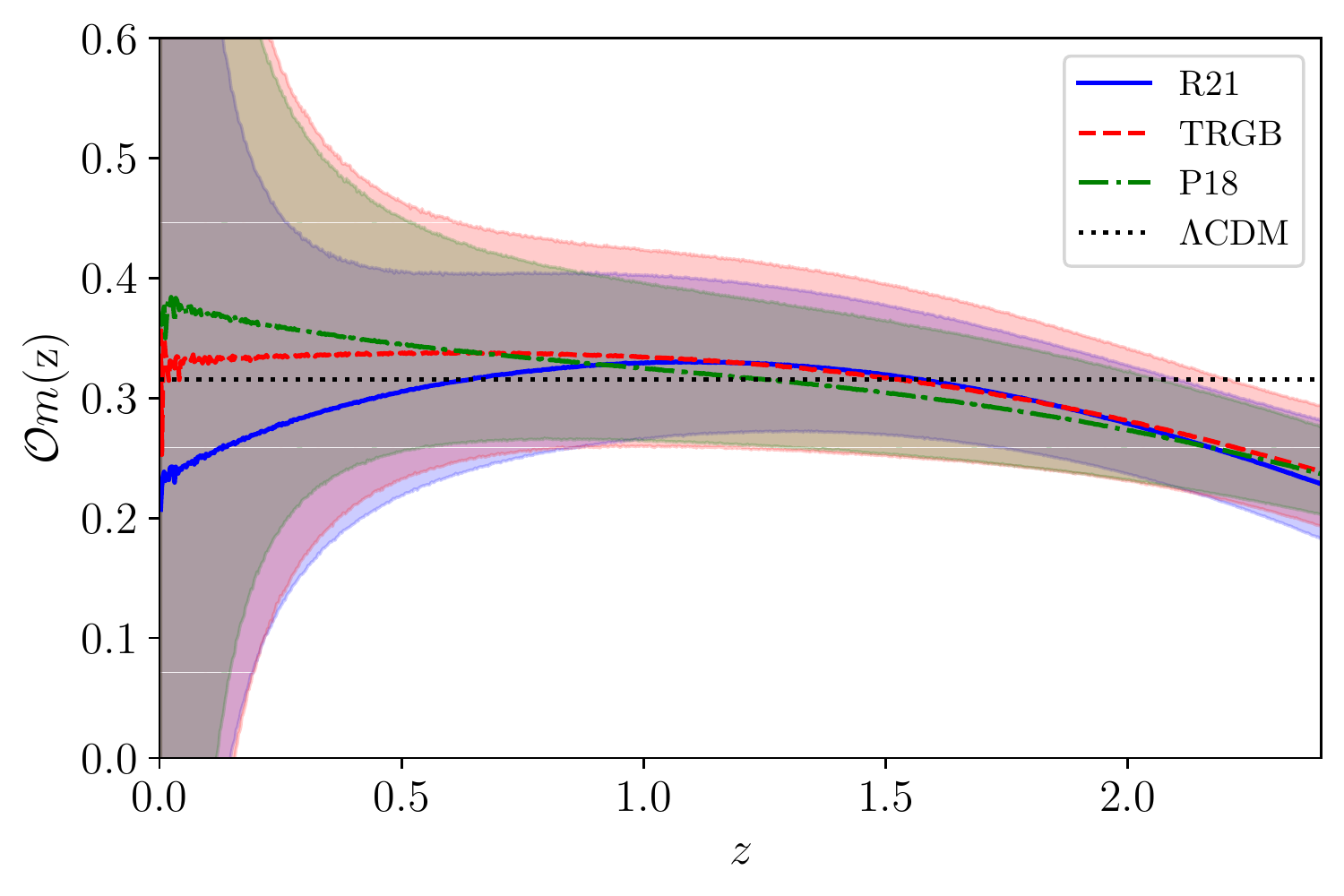}
\includegraphics[width=0.45\textwidth]{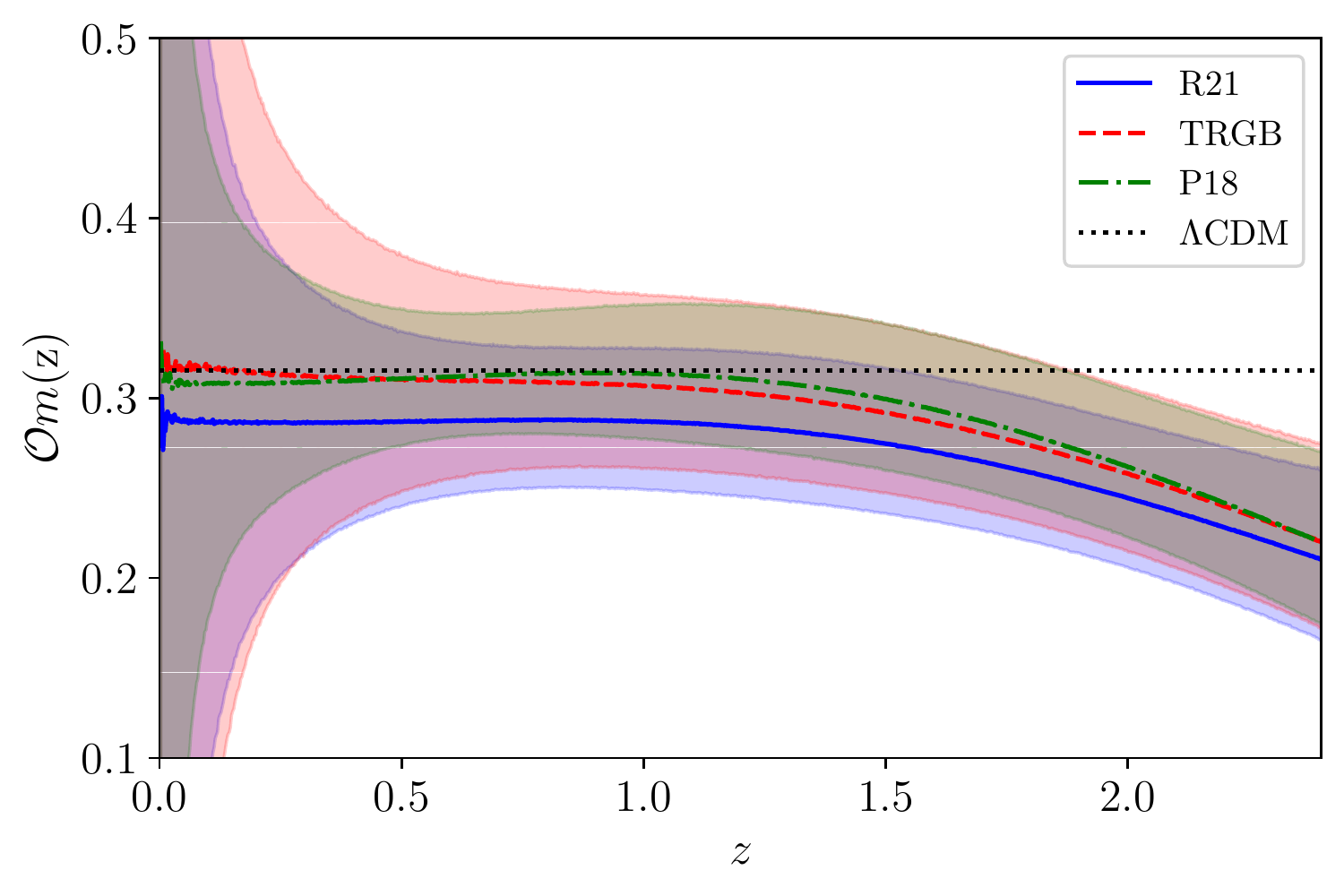}
\caption{Plots for the reconstructed $\mathcal{O}m$ diagnostics using (i) neural networks (left panel) and (ii) gaussian processes (right panel), from the combined CC+BAO Hubble data considering R21, TRGB, and P18 $H_0$ priors. }
\label{fig:om_diag_h}
\end{figure}

We plot the $\mathcal{O}m$ diagnostics, as a function of the redshift $z$, using the reconstructed $E(z)$ in Figs. \ref{fig:om_diag_sn} and \ref{fig:om_diag_h} from the Pantheon SN and combined CC+BAO Hubble data respectively.  The uncertainties associated with the reconstructed $\mathcal{O}m$ diagnostics are obtained by an MC error propagation technique. We also show a comparison between the two methods of reconstruction, i.e. implementation with neural networks in the left panel, and employing Gaussian processes in the right panel. Fig. \ref{fig:om_diag_sn} and \ref{fig:om_diag_h} show that the reconstructed values are not well constrained at lower redshifts $z < 0.2$. The mean reconstructed $\mathcal{O}m$ curves in both the figures show evolution with increasing redshift. In Fig. \ref{fig:om_diag_sn}, we find that the mean curves are characterised by a significant positive slope for $z > 1$, nonetheless the $\Lambda$CDM model assuming the Planck best-fit $\Omega_{m0}= 0.315$ \cite{Aghanim:2018eyx} is consistent with the $\mathcal{O}m$ reconstruction at the 2$\sigma$ confidence level. Whereas, the reconstruction profile in Fig. \ref{fig:om_diag_h} tends to be characterised by a negative slope for $z > 1$, excluding $\Lambda$CDM at 2$\sigma$ confidence level for $z>2$. This deviation from the concordance model possibly arises from the inclusion of high redshift Ly-$\alpha$ BAO measurements which calls for further investigation.

\subsection{\texorpdfstring{$H_0$}{} diagnostics} \label{sec:H0-diagnostics}

The Hubble tension, routinely presented as a mismatch between the Hubble constant $H_0$ determined from local measurements and a value inferred from the CMB sky assuming $\Lambda$CDM cosmology, essentially boils down to a disagreement between two numbers. Assuming this tension is cosmological in origin, the authors in \cite{Krishnan:2020vaf} explore the possibility of other inferred values of $H_0$, predicting that a ``running of $H_0$ with $z$'' may be expected within the concordance model. Similar possibilities of a steadily varying trend in the inferred $H_0$ as one moves from low to high redshift data have also been studied \cite{Krishnan:2022fzz, Colgain:2022nlb, Colgain:2022rxy, Colgain:2022tql, Dainotti:2021pqg, Dainotti:2022bzg, Schiavone:2022shz, Malekjani:2023dky}. Such a phenomenological evolution of $H_0$ with the $z$ could be a straightforward alternative in resolving the tension without any direct investigation of the fundamental framework. One such diagnostic that flags possible deviations from $\Lambda$CDM is the $H_0$ diagnostics $\mathbf{{H0}}$, defined as
\begin{equation}
    \mathbf{H0} = \frac{H(z)}{ \sqrt{\Omega_{m0}(1+z)^3 + 1-\Omega_{m0} }} \, . \label{eq:H0}
\end{equation}
This quantity $\mathbf{H0}$ provides us with a null test for the concordance model and a non-constancy of $\mathbf{H0}$ suggests evidence for new physics beyond $\Lambda$CDM. 

\begin{figure}[ht]
\centering
\includegraphics[width=0.45\textwidth]{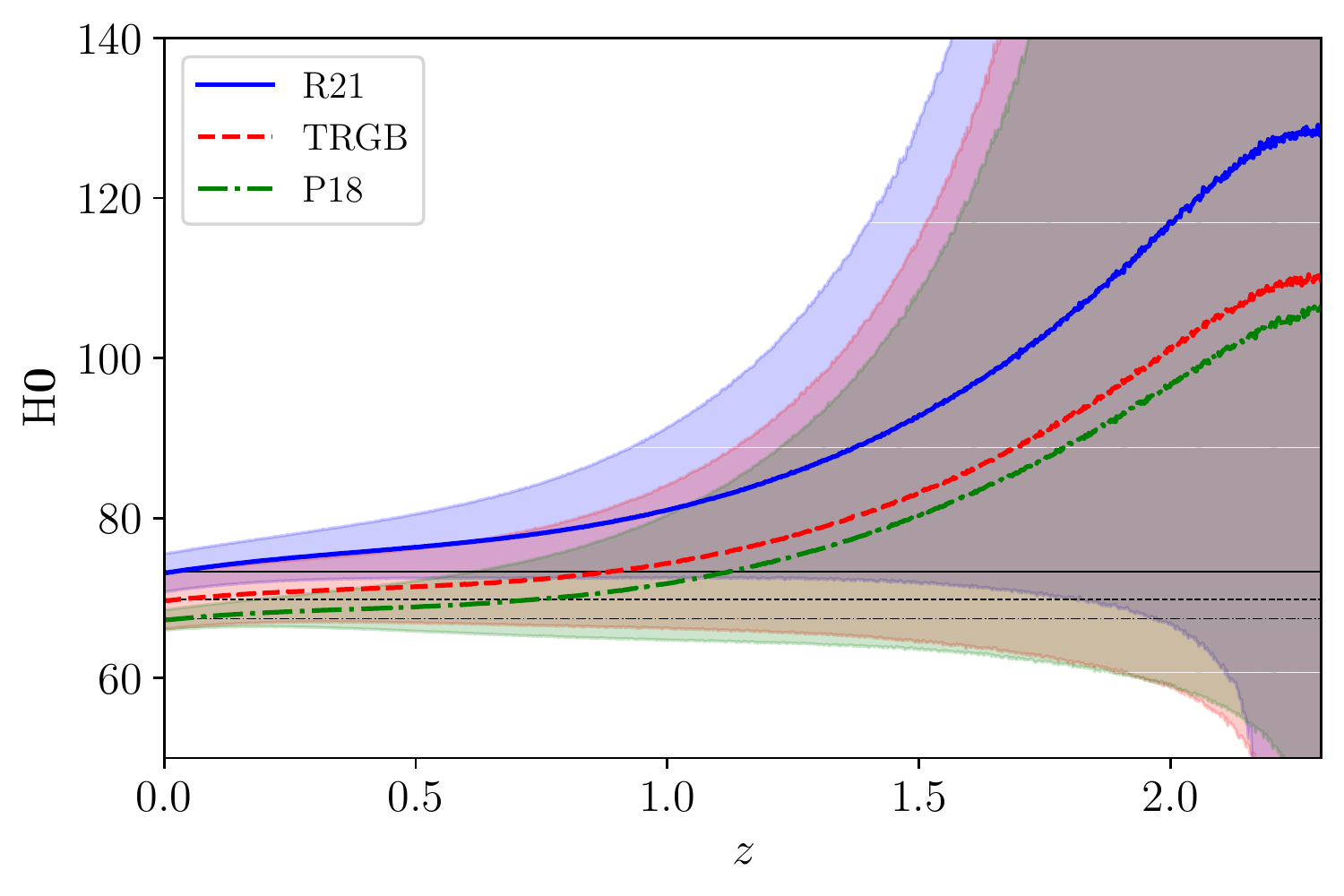}
\includegraphics[width=0.45\textwidth]{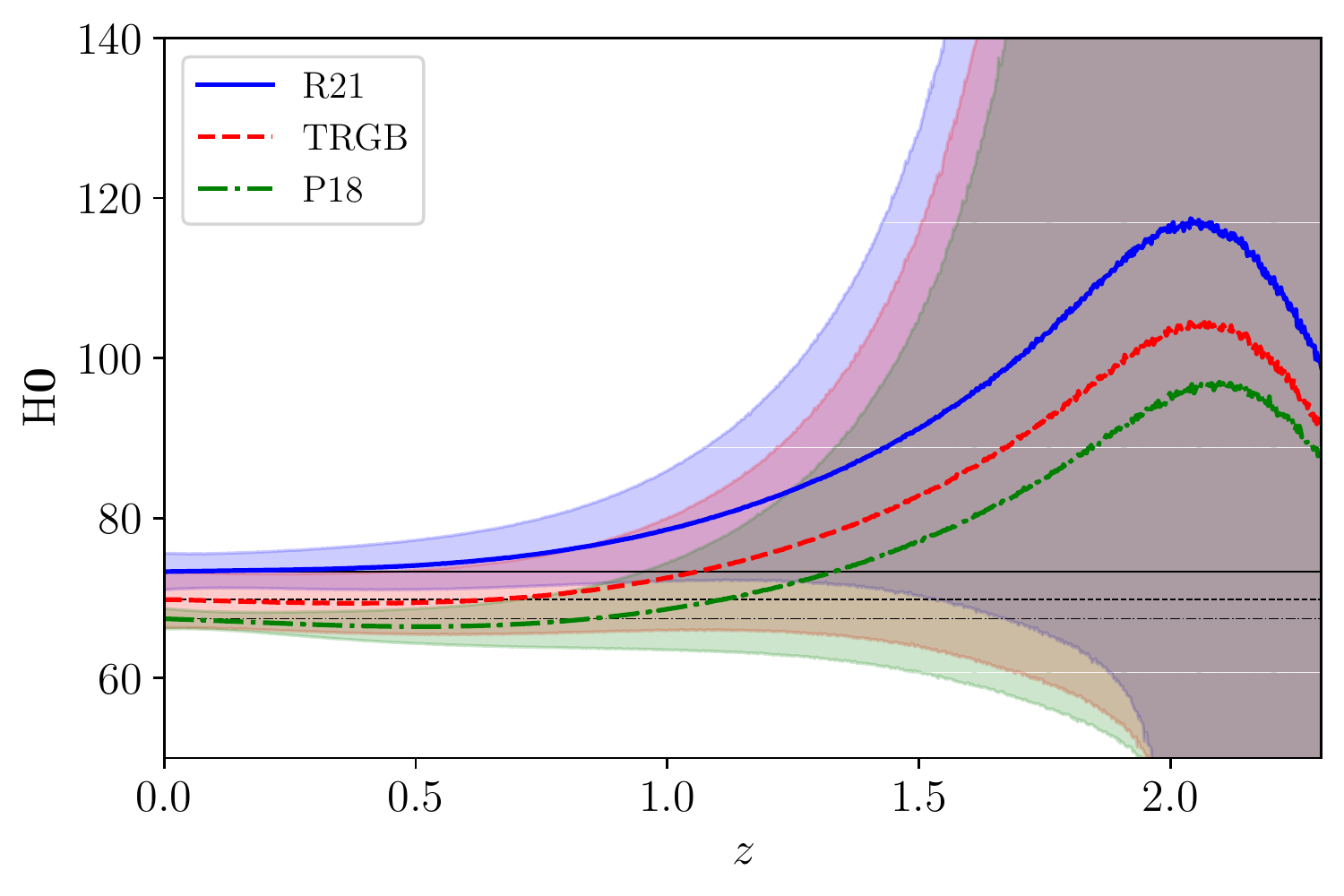}
\caption{Plots for the reconstructed $\mathrm{H_0}$ diagnostics using (i) neural networks (left panel) and (ii) gaussian processes (right panel), from the Pantheon SN data considering R21, TRGB, and P18 $H_0$ priors. }
\label{fig:H0_diag_sn}
\end{figure}

\begin{figure}[ht]
\centering
\includegraphics[width=0.45\textwidth]{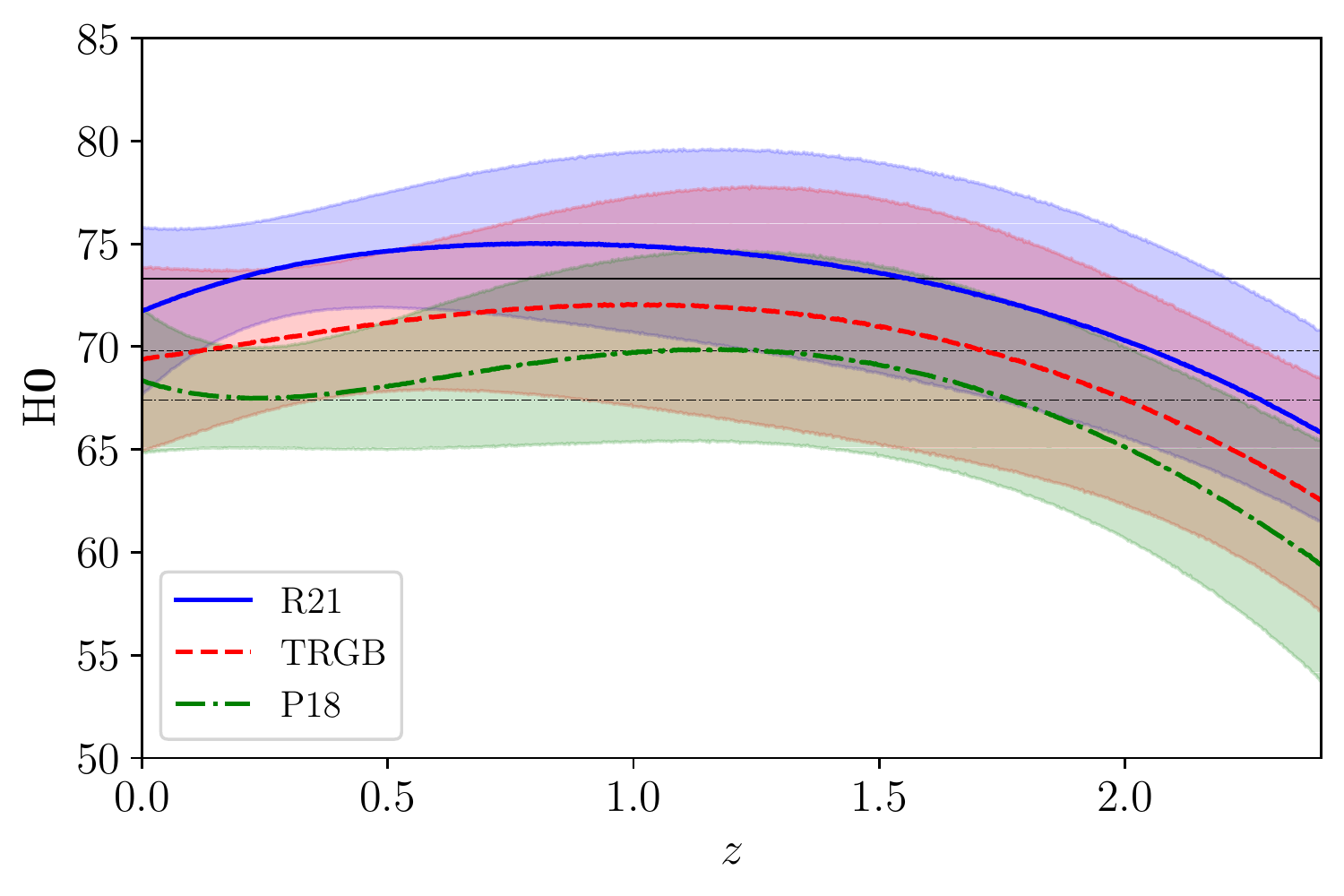}
\includegraphics[width=0.45\textwidth]{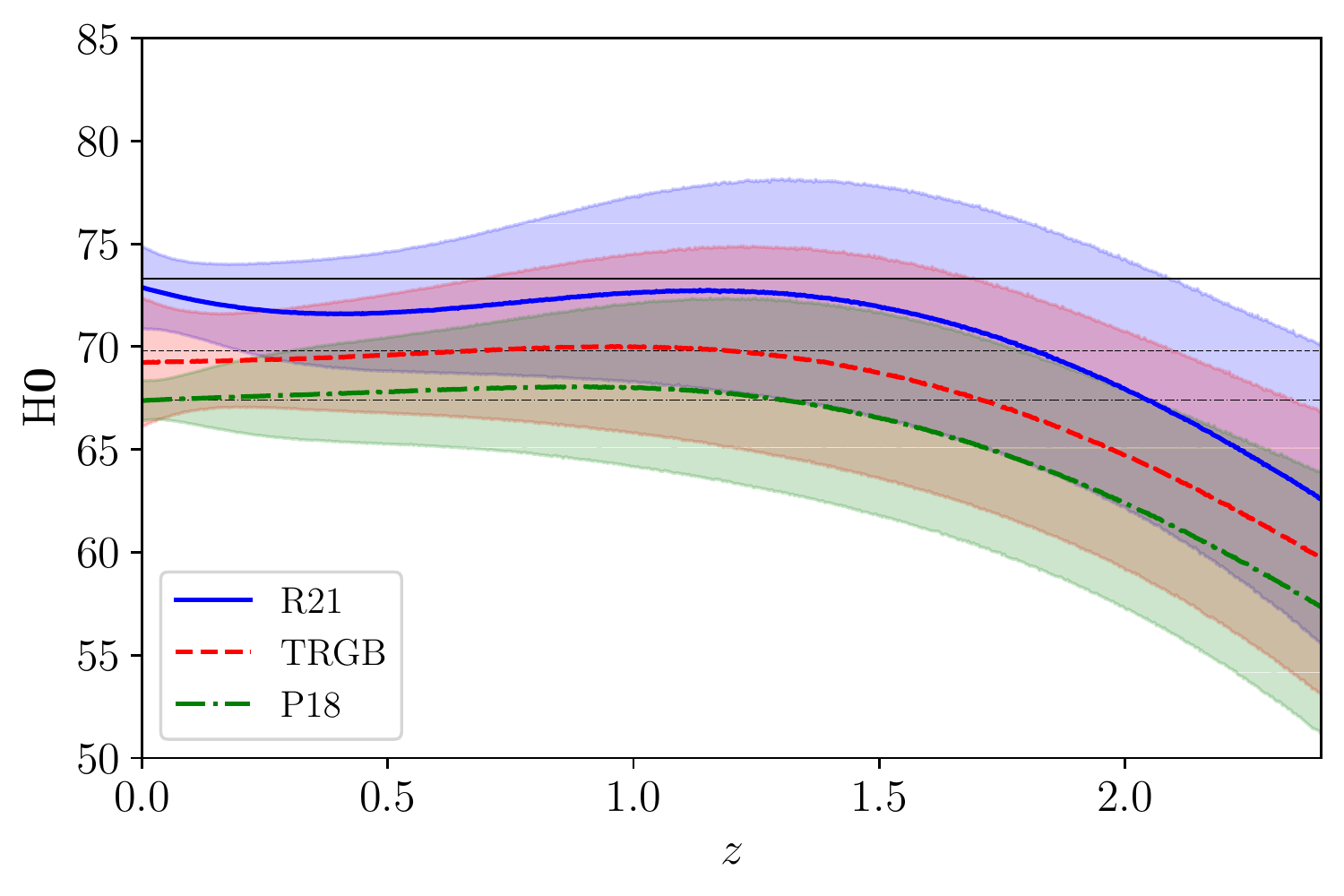}
\caption{Plots for the reconstructed $\mathrm{H_0}$ diagnostics using (i) neural networks (left panel) and (ii) gaussian processes (right panel), from the combined CC+BAO Hubble data considering R21, TRGB, and P18 $H_0$ priors. }
\label{fig:H0_diag_h}
\end{figure}

In this section, we plot the evolution of $\mathbf{H0}$ with respect to the redshift $z$ from the reconstructed $E(z)$ in Figs. \ref{fig:H0_diag_sn} and \ref{fig:H0_diag_h} from the Pantheon SN and combined CC+BAO Hubble data respectively. The left panels correspond to the reconstruction with ANNs, whereas the right panel represents the reconstruction using GPs. We make use of the employed $H_0$ priors to obtain the numerator $H(z) = H_0 E(z)$, in the RHS of \ref{eq:H0}. The denominator has been fixed by sampling $\Omega_{m0}$ directly via an MCMC analysis with the combined CC+BAO+SN data sets assuming $\Lambda$CDM cosmology. The constraints obtained on $\Omega_{m0}$ are $0.290 \pm 0.016$, $0.298 \pm 0.017$ and $0.303 \pm 0.016$ considering the R21, TRGB and P18 $H_0$ priors. The uncertainties associated with the parameter $\Omega_{m0}$ and reconstructed $H(z)$ are propagated using the MC error propagation technique.

Our results show that the mean reconstructed $\mathbf{H0}$ curves in both the figures show a non-monotonic evolution with respect to $z$. In Fig. \ref{fig:H0_diag_sn}, $\mathbf{H0}$ progressively increases with increasing $z$, but on going beyond $z>2$ we observe a dip in the reconstruction profile. The presence of such a dip is apparent in the right panel when employing GPs. We also plot the R21, TRGB, and P18 $H_0$ values in black solid, dashed and dotted lines to simultaneously compare them with the obtained $\mathbf{H0}(z)$ respectively. We find that the reconstructed errors accommodate $\Lambda$CDM within a $2\sigma$ level. The non-monotonic nature of $\mathbf{H0}$ is clearly visible in Fig. \ref{fig:H0_diag_h}, when the Hubble data is taken into consideration. The reconstructed $\mathbf{H0}$ profile indicates a clear deviation from $\Lambda$CDM at the 2$\sigma$ confidence level, driven by Lyman-$\alpha$ BAO leading to a significant dip in $\mathbf{H0}$ for $z>2$. However, if we restrict our attention to $z < 1$, where the quality of available data is much better, one finds little evidence for any deviation from $\Lambda$CDM cosmology.

\section{Conclusion \label{sec:conclusion}}

Even though reconstruction techniques have been a very popular topic of research the last few years in cosmology, the majority of the studies focus on GP to reconstruct dark energy and its potential theoretical foundations. GP, however, suffer from various problems among which are overfitting at low redshifts, meaning that the reconstructed function is too closely aligned to low redshift data points, as well as the selection of a kernel which introduces a statistical bias. 

ANNs have been proposed as a promising alternative to GPs, but in contrast to GPs, one can reconstruct only the cosmological parameters without their derivatives. There has been a recent work on the reconstruction of higher derivatives of the Hubble function in \cite{Mukherjee:2022yyq}, where the authors use an MC approach. Even though, this helps with the testing  of cosmological models, up to now there have been used only independent data points, while the most realistic data sets are correlated somehow. 

In this work, our goal was to include covariance information in the reconstruction approach in order to be able to use more realistic data sets. Once we reconstruct a cosmological parameter, we can use the Monte Carlo approach to reconstruct its higher derivatives and thus reproduce or test the viability of various cosmological models with better accuracy than before.

In greater detail, we reconstructed the Hubble diagram for various combinations of Cosmic Chronometers, Baryon Acoustic Oscillations, as well as the 1048 data points of Supernovae type Ia of Pantheon, which are correlated. To do this, we expanded \texttt{ReFANN}, that was initially formed based on \texttt{PyTorch}, using only independent data points. 

The type of data that ANNs can use is not as constrained as in GP. Specifically, ANNs make much less assumptions, because the many more hyperparameters they use, imitate in a better way the natural process compared to GP. For this reason, one would expect that, ANNs would produce higher uncertainties, however this is not the case here. Because of the large volume of data in the Pantheon set, both GP and ANNs perform in a similar way in terms of error bars. Thus, comparison between the two techniques shows more potential for the latter, since it does on exact training data and also can represent more complicated structures of data sets.

Last but not least, apart from the reconstruction of the Hubble function, we performed null tests in order to test the consistency of our results. In particular, through the $\mathcal{O}m$ and the $\mathbf{H0}$ diagnostics we tried to identify possible deviations from the $\Lambda$CDM model. Both diagnostics indicate a deviation from the concordance model at $z>2$, most probably because of the inclusion of the high redshift BAO data points. However, they both can accommodate $\Lambda$CDM at $2\sigma$ confidence level.

What would interesting to see from now on, is not only to forecast observations for experiments in progress that are about to publish their results, but also to use the reconstructed Hubble parameter and its derivative to constrain or even eliminate alternative cosmological models.

\begin{acknowledgements}

This paper is based upon work from COST Action CA21136 {\it Addressing observational tensions in cosmology with systematics and fundamental physics} (CosmoVerse) supported by COST (European Cooperation in Science and Technology). PM thanks ISI Kolkata for computational facilities and financial support through Research Associateship under project A/C No. 5756H. JLS and JM would like to acknowledge funding from ``The Malta Council for Science and Technology'' through the ``FUSION R\&I: Research Excellence Programme''. The work was supported by the PNRR-III-C9-2022–I9 call, with project number 760016/27.01.2023 and by the Hellenic Foundation for Research and Innovation (H.F.R.I.) under the ''First Call for H.F.R.I. Research Projects to support Faculty members and Researchers and the procurement of high-cost research equipment grant'' (Project Number: 2251).

\end{acknowledgements}


\bibliographystyle{spphys}
\bibliography{references}

\label{lastpage}

\end{document}